\documentstyle[12pt,epsf]{article}

\newlength{\pubnumber} 

\catcode`\@=11
\@addtoreset{equation}{section}

\def\section{\@startsection{section}{1}{\z@}{3.5ex plus 1ex minus .2ex}
 {2.3ex plus .2ex}{\large\bf}}
\def\subsection{\@startsection{subsection}{2}{\z@}{2.3ex plus .2ex}
 {2.3ex plus .2ex}{\bf}}

\def\bh{\bar{h}}
\def\bA{\bar{A}}
\def\bD{\bar{D}}
\def\bF{\bar{F}}
\def\bK{\bar{K}}
\def\bP{\bar{\Phi}}
\def\bS{\bar{S}}
\def\bX{\bar{X}}
\def\cD{{\cal{D}}}

\def\bPhi{\bar{\Phi}}
\def\P{\Phi}

\def\ss#1#2{\sum_{i=#1}^{#2} S_i \bS_i}
\def\sso#1#2{\sum_{i=#1}^{#2} S_{2i-1} S_{2i}}
\def\bsso#1#2{\sum_{i=#1}^{#2} \bS_{2i-1}\bS_{2i}}

\def\beq{\begin{equation}}
\def\eeq{\end{equation}}
\def\beqn{\begin{eqnarray}}
\def\eeqn{\end{eqnarray}}
\def\no{\noindent }
\def\nolabel{\nonumber }

\def\phl{\phantom{i}}

\def\gsim{{\buildrel >\over \sim}}

\def\half{{\textstyle{1\over 2}}}

\def\sixth{{\textstyle {1\over6}}}

\def\Tr{{\rm Tr}\, }

\def\vev#1{\langle #1\rangle}

\def\eps{\epsilon}

\def\inbar{\,\vrule height1.5ex width.4pt depth0pt}

\def\IC{\relax\hbox{$\inbar\kern-.3em{\rm C}$}}
\def\IQ{\relax\hbox{$\inbar\kern-.3em{\rm Q}$}}
\def\IR{\relax{\rm I\kern-.18em R}}
 \font\cmss=cmss10 \font\cmsss=cmss10 at 7pt

\def\IZ{\relax\ifmmode\mathchoice
 {\hbox{\cmss Z\kern-.4em Z}}{\hbox{\cmss Z\kern-.4em Z}}
 {\lower.9pt\hbox{\cmsss Z\kern-.4em Z}}
 {\lower1.2pt\hbox{\cmsss Z\kern-.4em Z}}\else{\cmss Z\kern-.4em Z}\fi}

\def\Io{\relax\ifmmode\mathchoice
 {\hbox{\cmss 1\kern-.4em 1}}{\hbox{\cmss 1\kern-.4em 1}}
 {\lower.9pt\hbox{\cmsss 1\kern-.4em 1}}
{\lower1.2pt\hbox{\cmsss 1\kern-.4em 1}}\else{\cmss 1\kern-.4em 1}\fi}

\def\be{\begin{equation}}
\def\ee{\end{equation}}
\def\bea{\begin{eqnarray}}
\def\eea{\end{eqnarray}}

\begin{document}
\begin{titlepage}
\setcounter{page}{1}
\rightline{BU-HEPP-05-08, CASPER-05-10}
\rightline{\tt hep-ph/0510141n}

\vspace{.06in}
\begin{center}
{\Large \bf Stringent Phenomenological Investigation into \\ 
            Heterotic String Optical Unification}
\vspace{.12in}

{\large
        J. Perkins,$^{1}$\footnote{john{\underline{\phantom{a}}}perkins@baylor.edu}
        B. Dundee,$^{1}$\footnote{ben{\underline{\phantom{a}}}dundee@baylor.edu}
        R. Obousy,$^{1}$\footnote{richard{\underline{\phantom{a}}}k{\underline{\phantom{a}}}
                        obousy@baylor.edu}
        S. Hatten, $^{1,2}$\footnote{hattst@wwc.edu}
        E. Kasper, $^{1,3}$\footnote{ekasper@physics.tamu.edu}
        M. Robinson, $^{1,4}$\footnote{m{\underline{\phantom{a}}}robinson@baylor.edu}
        C. Sloan, $^{1,5}$\footnote{cwsloan@edisto.cofc.edu}
        K. Stone, $^{1,6}$\footnote{zkrs18@imail.etsu.edu}
        and G. Cleaver,$^{1}$\footnote{gerald{\underline{\phantom{a}}}cleaver@baylor.edu}}
\\
\vspace{.12in}
{\it $^{1}$ Center for Astrophysics, Space Physics \& Engineering Research\\
            Department of Physics, Baylor University,
            Waco, TX 76798-7316\\}
\vspace{.06in}
{\it $^{2}$ Walla Walla College, 204 South College Avenue\\
            College Place, WA 99324\\}
{\it$^{3}$  Dept. of Physics, Texas A \& M University,\\ 
            College Station, TX 77843-4242\\} 
\vspace{.06in}
{\it$^{4}$  Dept. of Physics, Auburn University,\\ 
            Auburn, AL 36849\\}
\vspace{.06in}
{\it$^{5}$Dept. of Physics, College of Charleston\\
Charleston, SC 29424\\}
\vspace{.06in}
{\it$^{6}$  Dept. of Physics, East Tennessee State University\\
            Johnson City, TN 37614}
\end{center}

\begin{abstract}

For the weakly coupled heterotic string (WCHS)
there is a well-known factor of twenty conflict 
between the minimum string coupling unification scale, $\Lambda_H \sim 5 \times 10^{17}$ GeV,
and the projected MSSM gauge coupling unification scale,
$\Lambda_U \sim 2.5 \times 10^{16}$ GeV,
assuming an intermediate scale desert (ISD).
From a bottom-up approach, renormalization effects of intermediate scale 
MSSM-charged exotics (ISME), which are endemic to quasi-realistic string models,
can resolve this issue by pushing the
MSSM scale up to the string scale. However, for a generic string model, 
this implies that the projected $\Lambda_U$ unification under the ISD assumption 
is accidental. 

If the true unification scale
is $\Lambda_H\gsim 5.0 \times 10^{17}$ GeV, is it possible that an illusionary unification at 
$\Lambda_U = 2.5 \times 10^{17}$ GeV in the ISD scenario is not accidental? This is an issue recently raised again 
by Bin\' etruy et al.\ \cite{tfq}.
If it is not accidental, then under what conditions
would the assumption of ISME in a WCHS model imply an apparent unification at 
$\Lambda_U < \Lambda_H$ when an ISD is falsely assumed? 
J.\ Geidt's {\it optical unification} suggests that $\Lambda_U$  
is not accidental and provides sufficient conditions for the appearance of $\Lambda_U$.
In fact, through constrained ISME, optical unification offers a mechanism whereby 
a generic MSSM scale $\Lambda_U < \Lambda_H$ is guaranteed.  

A WCHS model was recently constructed 
that offers the possibility of optical unification \cite{optun1}. Whether optical unification
can be realized depends on the availability of anomaly-cancelling $D$- and $F$-flat directions
meeting certain phenomenological requirements \cite{optun2}.
This paper reports on the systematic investigation of the optical unification properties
of a subset of flat directions of this model that are {\it stringently} flat. Stringent flat directions
can be guaranteed to be $F$-flat to all finite order (or to at least a given finite order consistent with
electroweak scale supersymmetry breaking) and can be viewed as the likely roots of more general flat directions.
Analysis of the phenomenology of stringent flat directions gives an indication of the remaining optical unification 
phenomenology that must be garnered by flat directions developed from them.\\
{\it This paper is a result of the 2003-2004 NSF REU program at Baylor University.}
\end{abstract}
\end{titlepage}
\setcounter{footnote}{0}

\section{Review of Optical Unification}

The lower limit to string
coupling unification in a weakly coupled heterotic string (WCHS)
was shown by Kaplunovsky in 1992 to be around
$\Lambda_H \sim 5 \times 10^{17}$ GeV \cite{kapl}.
In contrast, under the scenario of an intermediate scale desert (ISD),
the runnings of the $SU(3)_C\times SU(2)_L\times U(1)_Y$ ([321]) couplings in
the Minimal Supersymmetric Standard Model (MSSM) predict
a unification scale $\Lambda_U \sim 2.5 \times 10^{16}$ GeV \cite{mssmisd}. 
The issue of this factor-of-twenty difference was raised again in the third of 
the {\it Twenty-Five Questions for String Theorists} by  
Bin\' etruy et al.\  \cite{tfq}.

One resolution to the factor-of-twenty difference 
between these two scales.  
is a grand unified theory (GUT) 
between the MSSM and string scales. However,
with the exception of flipped $SU(5)$ \cite{fsu5}
(or partial GUTs such as the Pati-Salam $SU(4)_C\times SU(2)_L\times SU(2)_R$
\cite{lrs,sguts},
string GUTs cannot be generated by level-one Ka\v c-Moody algebras 
(since they lack the required 
adjoint Higgs and/or higher dimensional scalar representations) 
and models based on higher level Ka\v c-Moody algebras vastly prefer even 
numbers of generations \cite{hlguta,hlgutb,hlgutc}. 
Alternately, strong coupling effects of
$M$-theory can lower $\Lambda_H$ down to $\Lambda_U$ \cite{witten}.
On the other hand, 
intermediate scale MSSM-charged exotics (ISME)
at $\Lambda_I< \Lambda_U$ could shift
the MSSM unification scale upward to the string scale \cite{mup}.
The near ubiquitous appearance of ISME in (quasi)-realistic 
heterotic string semi-GUT \cite{psm2,sguts}, (near)-MSSM \cite{nmssm,af3,mssm1,mssm2}
and GUT \cite{fsu5} models adds weight to the third proposal.  
However, most intermediate scale MSSM-charged 
exotic solutions might be viewed as accidental.  

While existence of 
the string unification scale would clearly be stable under shifts 
in the masses of the exotics, the prediction of an apparent MSSM unification
scale when MSSM-charged exotics are ignored would generally by unstable under 
these mass shifts.  
On the other hand, a set of ISME satisfying {\it optical unification} constraints provide
a robust method for stabilizing an apparent MSSM unification scale 
under such shifts \cite{jg1}. 
In optical unification, ISME 
affect running couplings like a diverging lens, always producing a ``virtual'' 
image of the string unification point between the string scale and the
exotic particle mass scale. 
That is, a shift of the intermediate scale $\Lambda_I$ simply
produces a shift in $\Lambda_U$, rather than the disappearance of $\Lambda_U$.
Thus, a string model with optical unification
offers a resolution to question three of \cite{tfq}.

Successful optical unification requires three things \cite{jg1}.
First, the effective level of the
hypercharge generator must be the standard
\beqn
k_Y = \textstyle{5\over3}.
\label{db123k}
\eeqn
(\ref{db123k}) is a
strong constraint on string-derived $[321]$ models,
for the vast majority have non-standard
hypercharge levels.
Only select classes of models, such as the 
free fermionic \cite{fff} NAHE-based \cite{nahe} class,
can yield $k_{Y} = \textstyle{5\over3}$.

Second, optical unification imposes the relationship
\beqn
\delta b_2 = \textstyle{7\over12} \delta b_3 + \textstyle{1\over4} \delta b_Y.
\label{db123}
\eeqn
between the exotic particle contributions $\delta b_3$, $\delta b_2$, 
and $\delta b_1$
to the [321] beta-function coefficients.
Each $SU(3)_C$ exotic triplet or anti-triplet contributes
$\half$ to $\delta b_3$;
each $SU(2)_C$ exotic doublet contributes
$\half$ to $\delta b_2$.
With the hypercharge of a MSSM quark doublet normalized to $\sixth$,
the contribution to $\delta b_Y$ from an individual particle with
hypercharge $Q_Y$ is $Q_Y^2$.
$\delta b_3 > \delta b_2$ is required
to keep the virtual unification scale below the string scale.
In combination with (\ref{db123}), this imposes
\beqn
\delta b_3 >  \delta b_2 \ge \textstyle{7\over12} \delta b_3,
\label{db123b}
\eeqn
since $\delta b_Y \geq 0$.

To acquire intermediate scale mass,
the exotic triplets and anti-triplets must be equal in number.
Similarly, an even number of exotic doublets is required.
Hence, $\delta b_3$ and $\delta b_2$ must be integer.
The simplest solution to
(\ref{db123}) and (\ref{db123b}) is a set of
three exotic triplet/anti-triplet pairs and two pairs of doublets.
One pair of doublets can carry $Q_Y=\pm \half$, while the remaining
exotics carry no hypercharge \cite{jg1}.
Alternately, if the doublets carry too little hypercharge,
some exotic $SU(3)_C \times SU(2)_L$ singlets could make up the
hypercharge deficit.
The next simplest solution requires four triplet/anti-triplet pairs and three pairs of
doublets that yield $\delta b_Y = 2 \textstyle{2\over 3}$
either as a set, or with the assistance of additional non-Abelian singlets.
For models containing more than four triplet/anti-triplet pairs, (\ref{db123})
and (\ref{db123b}) allow varying numbers of pairs of doublets.

In Section 2 we review the particle content of the optical unification model. In Section 3
we review flatness constraints of the WCHS and discuss the properties that optical unification 
flat directions must possess. In Section 4 we present the findings of our investigation 
of stringent flat directions for optical unification. These results are then summarized in Section 5.

\section{Heterotic String Model with Optical Unification Potential} 

A search for free fermionic WCHS models with the potential for 
optical unification was recently conducted \cite{optun1}.  
One such model (see Tables A.1-4) was discovered by altering GSO projection coefficients 
of a model in \cite{af3}.
From this, a new model was constructed that contains 
a set of 4 $SU(3)_C$ exotic triplet/anti-triplet pairs, 3 $SU(2)_L$ exotic doublets, 
and a pair of non-Abelian singlets (chosen from a set of seven such pairs) 
that together satisfy optical unification requirements \cite{optun1,optun2} (see Table A.2). 
Three pairs of exotic triplet/anti-triplets carry hypercharge $Q_Y= \pm \frac{1}{3}$, 
while one pair carries $Q_Y= \pm \frac{1}{6}$. 
All three pairs of exotic doublets carry $Q_Y = 0$ while the 
pairs of non-Abelian singlets carries $Q_Y=\pm \frac{1}{2}$. The only additional exotic MSSM
(besides the above mentioned six extra pairs of singlets) are the 
three extra pairs of MSSM Higgs doublets endemic to (quasi)-realistic heterotic models (see Table A.2). 

The optical unification constraints (\ref{db123},\ref{db123b})
require that, together, the four triplet/anti-triplet pairs, the three exotic
doublet pairs, and exactly one of the pairs of hypercharged exotic singlets
form the set of ISME at $\Lambda_I < \Lambda_U$. Thus, 
the remaining six pairs of exotic hypercharged singlets and the three extra Higgs 
must take on $\Lambda_U$ scale (or higher) masses. 

Like most quasi-realistic heterotic string models 
the possible optical unification model contains an anomalous $U(1)_A$
(i.e., for which ${\rm Tr}\, Q^{(A)}\ne 0$) \cite{anomu1}. For this model
\beqn
{\rm Tr}\, Q^{(A)} = +72,
\eeqn
with a net contribution of +24 from the standard MSSM three generations, of +48 from the
hidden sector non-Abelian matter states, and no net contribution from the exotic non-Abelian 
singlets. 
 
The set of MSSM-uncharged matter states is composed of
27 non-Abelian singlet fields, henceforth denoted ``singlets'' (see Table A.3), 
and 16 hidden sector non-Abelian fields (see Table A.4). 
The singlet fields are
$\Phi_{i=1,2,3}$ (the three totally uncharged moduli), 
$\Phi_{12}$, $\Phi_{23}$, $\Phi_{31}$ and complex conjugate fields
$\bar{\Phi}_{12}$, $\bar{\Phi}_{23}$, $\bar{\Phi}_{31}$,
and $S_{j=1\, {\rm to }\, 9}$ and complex conjugate fields $\bar{S}_{j}$. 
Except for the three uncharged moduli, all singlets form vector-like pairs, 
$(\Phi_{ij},\bar{\Phi}_{ij})$ and $(S,\bar{S})_k$. 
Of these, only $S_{7}$, $S_{8}$, and $S_{9}$ (and $\bar{S}_{7}$, $\bar{S}_{8}$, 
and $\bar{S}_{9}$) carry anomalous charge, which is positive for $S_{7}$, $S_{8}$, and $\bar{S}_{9}$ 
and negative for their vector partners. 

The set of hidden sector non-Abelian states is composed of 
(i) four $SU(5)_H$ $\mathbf 5$ reps, $F_{1,\, 2,\, 3,\, 4}$, and
    four $\bar{\mathbf 5}$ reps, $\bar{F}_{1}$, 
$\bar{F}^{'}_{2,\, 3,\, 4}$, 
and (ii) four $SU(3)_H$ $\mathbf 3$ reps, $K_{1,\, 2,\, 3,\, 4}$, 
and four $\bar{\mathbf 3}$ reps,  
$\bar{K}^{'}_{1,\, 2,\, 3}$, $\bar{K}_{4}$. 
$(F_{1},\bar{F}_{1})$ and $(K_{4},\bar{K}_{4})$  
form vector-like pairs of states, while $'$ indicates
$F_{n}$ and $\bar{F}^{'}_{n}$ and $K_{n}$ and $\bar{K}^{'}_{n}$ do not form vector-like pairs, 
but, instead, have some matching charges. 

\section{Heterotic String Flat Directions} 

For heterotic strings,
the Green-Schwarz-Dine-Seiberg-Wittten mechanism \cite{gsdsw} breaks
the anomalous $U(1)_A$, and in the process generates 
a Fayet-Iliopoulos (FI) term,
\beqn
\eps\equiv\frac{g^2_s M_P^2}{192\pi^2}\Tr Q^{(A)},
\label{fit}
\eeqn
in the associated $D$-term.
The FI term breaks supersymmetry near the Planck scale 
and destabilize the string vacuum, unless it is cancelled 
by scalar vacuum expectation values (VEVs),
\beqn
\vev{D_{A}} &=& \sum_m Q^{(A)}_m |\vev{\phi_{m}}|^2 +
\eps  = 0\,\, . \label{daf}
\eeqn
Thus, an anomalous $U(1)_A$ induces 
a non-perturbatively chosen flat direction of VEVs. 
Since the fields taking on the VEVs
typically carry additional non-anomalous charges, a
non-trivial set of constraints is imposed on the VEVs.
The VEVs must maintain $D$-flatness for each non-anomalous gauge symmetry. 
For Abelian gauges, 
\beqn
\vev{D_i} &=& \sum_m Q^{(i)}_m |\vev{\phi_{m}}|^2 = 0\,\, ,
\label{dana}
\eeqn
while for non-Abelian gauges, 
\beqn
\vev{D_a^{\alpha}}&=& \sum_m
\vev{\phi_{m}^{\dagger} T^{\alpha}_a \phi_m} = 0\,\, ,
\label{dtgen}
\eeqn
where $T^{\alpha}_a$ is the $\alpha^{\rm th}$ matrix generator for scalar state $\phi_{m}$ in the 
representation $R$ of the gauge group ${\cal{G}}_a$. 
Since the states with anomalous charge often carry additional, non-anomalous charge,
their VEVs will in general break some, or all, of the non-anomalous gauge
symmetries spontaneously.

If matrix generators are ${T^{\alpha}_a}$ for states in the representation $R$, 
then the matrix generators are
\beqn
\bar{T}^{\alpha}_a = - {T^{\alpha}_a}^{\ast}. 
\label{barT}
\eeqn
for states in the representation $\bar{R}$.
Thus, for $SU(n)$ groups, the non-Abelian $D$-term 
contributions for vector-like pairs of non-Abelian states 
can cancel out.

To insure a supersymmetric vacuum, $F$-flatness,
\beq
\vev{F_{m}} \equiv \vev{\frac{\partial W}{\partial \Phi_{m}}} = 0,
\label{ff}
\eeq
must also be maintained for each superfield $\Phi_{m}$ (containing a scalar field $\phi_{m}$
and chiral spin-$\half$ superpartner $\psi_m$) appearing
in the superpotential $W$ (for which flatness is also required, i.e.\ $\vev{W} = 0$).

Optical unification places strong constraints on viable flat directions for this model.
A good optical unification flat direction must, as discussed,
\begin{itemize}
\item keep all of the MSSM exotic triplets (the D's) and doublets (the X's) massless 
above the intermediate scale $\Lambda_I$, where the optical-unification-producing
diverging lense effect occurs,  
\item generate $\Lambda_U$ or greater scale mass for six out of seven pairs of the exotic singlets (the A's), 
but keep one pair of exotic singlets massless down to the $\Lambda_I$ scale, and
\item generate $\Lambda_U$ or greater masses for three out of four pairs of the MSSM Higgs. 
\end{itemize}
Possible mass terms through seventh order in the superpotential for the exotic triplets, exotic doublets,
exotic singlets, and MSSM Higgs are given in  
Tables C.1a,b, Tables C.2a,b, Tables C.3a,b, and Tables C.4a,b of Appendix C, respectively.  

In WCHSs, FI term 
cancellation generically imposes 
scalar VEVs, $<\phi>$, of order .01 $M_{\rm Pl} \sim 1.2 \times 10^{17}$ Gev,
which is approximately $0.3 \Lambda_{H}$. For this model, the average $D$-flat 
direction that is also $F$-flat to at least 6$^{\rm th}$ order has an anomalous charge of around 12.
This corresponds to an average FI VEV scale of 
\beqn
|<\phi>|^2 &\sim& \eps/Q^{(A)}_{\phi}\equiv\frac{g^2_s M_P^2}{192\pi^2} (\Tr Q^{(A)}/Q^{(A)}_{\phi}),
\label{fivev}\\
           &=& (0.75 \times 1.2\times 10^{19}\, {\rm Gev})^2/(192 \pi^2) (72/12)
\nolabel\\
           &=&  5\times 10^{17} \sim \Lambda_{H}.
\eeqn
(taking $g_s\sim 0.75$). 
For free fermionic WCHSs, worldsheet charge constraints limit
dimensionless couplings, $\lambda_3$, 
in the third order superpotential to discrete values of 
$1/(\sqrt{2})^{i}$, for $i \in \{0,\, 1,\, 2\}$. 
Thus, masses from third order terms,
\beqn
\lambda_n <\phi> \bar{\Phi} \Phi,
\label{rwn}
\eeqn
are on the order of the string scale $\Lambda_{H}$. They are, thus, greater than the MSSM unification scale,
$\Lambda_{U}$, by a factor of 20 or more.

On the other hand, non-renormalizable terms of order $n > 3$, 
\beqn
\lambda_n \bar{\Phi}{\Phi} <\phi> \left(\frac{<\phi>}{M_{\rm Pl}}\right)^{n-3} ,
\label{nrwn}
\eeqn 
produce mass suppression. Factors of order $(1/100)^{n-3}$ are acquired 
from $(<\phi>/M_{\rm Pl})^{n-3}$. However, these are partially counter-acted by
the worldsheet phase space factor in $\lambda_n$, $n>3$. This yields a net mass suppression
factor of around (1/10), per increase in superpotential order. 
As pointed out in \cite{msup},
mass suppression factors actually begin at fifth, rather than fourth, order. 
At fourth order, the dimensionless coupling $\lambda_4$ can take on values 
as large as 10 to 100, due to integration over worldsheet phase space.    
Thus, mass terms from fourth order superpotential terms 
need not yield suppression, but can be on par with (or larger than) masses from third order 
superpotential terms. Hence, factor of (1/10) suppression per order
begins at fifth order. 

Therefore, masses from third through fifth or sixth order 
superpotential terms are above (or on par with) the MSSM unification scale, $\Lambda_U$, for
$\Lambda_H \sim 5\times 10^{17}$ GeV. 
For a somewhat higher WCHS scale, seventh order mass terms
might also be viable. 

\section{Optical Unification Investigation}

A typical WCHS model contains 
a moduli space of perturbative solutions to the $D$- and $F$-flatness constraints,
which are supersymmetric and degenerate in energy \cite{moduli}. 
Study of the phenomenology of superstring models often
involves the analysis and classification of these flat directions.
Thus, methods for flat direction analysis have been systematized
in recent years \cite{system,mshsm,hetdisc}.
Since our optical unification model contains an anomalous $U(1)_A$,
some of the scalar fields will necessarily receive FI-scale VEVs 
to cancel the FI term. (We assume for obvious reasons that 
the MSSM-charged scalars do not receive a VEV.) 

In this section we report on our investigation of $D$- and $F$-flat directions for our
optical unification model.
In general, systematic analysis of simultaneously $D$- and $F$-flat directions is a complicated, 
very non-linear process. In WCHS model-building $F$-flatness 
of a specific VEV direction in the low energy effective field theory may be proven to a given order 
(by cancellation of $F$-term components),
only to be lost a mere one order higher.  

To systematize analysis of $F$-flat direction, the all-order stringent approach was 
developed \cite{mshsm,hetdisc,sguts}. 
Rather than allowing cancellation between two or more components in an  
$F$-term, stringent $F$-flatness requires that each possible component in an $F$-term have zero 
vacuum expectation value.
When only non-Abelian singlet fields acquire VEVs, this implies that 
two or more singlet fields in a given $F$-term cannot take on VEVs. 
This condition can be relaxed when non-Abelian fields  
acquire VEVs. Self-cancellation of a single 
component in a given $F$-term is possible between various VEVs within non-Abelian reps.
Self-cancellation was discussed in \cite{mssm2,geoflat} for $SU(2)$ and $SO(2n)$ states. 
In the optical unification model, 
self-cancellation is possible through VEVs of $SU(3)$ ${\mathbf 3}$ and $\bar{\mathbf 3}$ reps
and also through VEVs of $SU(5)$  ${\mathbf 5}$ and $\bar{\mathbf 5}$ reps.
The $SU(n)$ self-cancelling VEV combinations are extensions of the 
$SU(2)$ examples presented in \cite{mssm2,geoflat}.

At least three $\mathbf 3$ and $\bar{\mathbf 3}$  
fields that must receive VEVs for $SU(3)$ self-cancellation. 
In the minimal case, two VEV fields must be $\mathbf 3$'s and one VEV field must be 
a $\bar{\mathbf 3}$ (or vice versa). 
In this case, if triplets
${\mathbf 3}_i$, $i= 1$, 2 receive respective VEVs
\beqn 
\exp\{i \pi \theta_i \} <R_i,\, G_i,\, 0>,
\label{rgbi}
\eeqn
where $R_i$ and $G_i$ are respective magnitudes (up to a sign) of hidden red and green charges,  
and $\exp\{i \pi \theta_i \}$ are respective overall phases,
then $SU(3)$ $D$-flatness  
is maintained by a $\bar{\mathbf 3}$ with VEV    
\beqn
\exp\{i \pi \bar{\theta} \} 
<\bar{R} = \sqrt{{R_1}^2 + {R_2}^2},\,  \bar{G} = \sqrt{{G_1}^2 + {G_2}^2},\, 0>.
\label{brgb}
\eeqn
Then, clearly in an $F$-term containing 
\beqn
\epsilon_{\alpha \beta \gamma} {\mathbf 3}_1^{\alpha} \cdot {\mathbf 3}_2^{\beta} 
\label{scab}
\eeqn
(with $\alpha$,  $\beta$, and $\gamma$ color indices),
self-cancellation occurs if
\beqn 
R_1 G_2 = G_1 R_2.  
\label{crbg}
\eeqn
Note that (\ref{crbg}) implies 
\beqn
R_1/G_1 = R_2/G_2 = \bar{R}/\bar{G}.
\label{crbg2}
\eeqn

The next simplest $SU(3)$ self-cancellation can occur between two triplet VEVs and two anti-triplet 
VEVs.\footnote{From gauge invariance arguments, it can be shown that triplets from a $D$-flat antisymmetrized 
$\epsilon_{\alpha \beta \gamma}  
{{\mathbf 3}_1}^{\alpha} 
{{\mathbf 3}_2}^{\beta} 
{{\mathbf 3}_3}^{\gamma}$ 
combination cannot produce self-cancellation.}
$D$-flatness is maintained by
\beqn
&&
<    {\mathbf 3}_1> = \exp\{i \pi \theta_1 \} <R,\ 0,\, 0>,\, 
<\bar{\mathbf 3}_1> = \exp\{i \pi \bar{\theta}_1 \}<\bar{R} = R,\ 0,\, 0>,
\label{ex24a}\\
&&
<    {\mathbf 3}_2> =  \exp\{i \pi \theta_2 \} < 0,\ G,\, 0>,\, 
<\bar{\mathbf 3}_2> =  \exp\{i \pi \bar{\theta}_2 \} < 0,\, \bar{G} = G,\ 0 >. 
\label{ex24b}\
\eeqn
Self-cancellation then occurs in $F$-terms containing either
\beqn
{\mathbf 3}_1\cdot \bar{\mathbf 3}_2,\,\, {\rm or}\,\,
 {\mathbf 3}_2\cdot \bar{\mathbf 3}_1.
\label{sc24b}
\eeqn 
These two self-cancellation classes can be generalized for more field VEVs.

\subsection{$D$-flat Basis Directions}

Our first step in a systematic search for optical unification producing $D$- and $F$-flat directions   
was to construct a basis of $D$-flat directions for the set of singlet fields with null hypercharge,
and for the set of hidden sector non-Abelian fields.  
We generated a set of 24 $D$-flat directions
$\{ \cD_i,\,\, {\rm for}\,\, i\, = \, 1,\,\, {\rm to}\,\, 24\}$ (see Tables 4.1 and 4.2) 
en mass via the singular value decomposition approach (described in \cite{svg}
and applied in \cite{mshsm}).  
Note that by {\it basis} of $D$-flat directions,
we mean a basis of directions specifically $D$-flat with regard to all
 {\it non-anomalous} $U(1)$ gauge symmetries. The $D$-flat basis elements may carry positive, 
negative, or zero anomalous charge. 
For a linear combination of basis directions to be {\it physical}, its 
net anomalous charge must be of opposite sign to the FI term. Thus, in this model its 
net anomalous charge must be negative. 
 
\vspace{0.2cm}
\no Table 4.1: $D$-Flat Singlet VEV Basis Elements.
\vspace{0.1cm}
\begin{flushleft}
\begin{tabular}{|l|r|lr|rrrrrr|}
\hline
Dir& ${Q^{(A)}}'$& identifying field &   & $S_8$&$\bP_{31}$& $S_2$& $S_4$ & $S_5$ & $S_6$\\
\hline
${\cal D}_1$& -1& $S_9      = $& 3 &    -3&  2&  0&  0& -1& -1\\
${\cal D}_2$&  0& $S_7      = $& 1 &    -1&  1&  0&  0& -1& -1\\
${\cal D}_3$&  0& $\bP_{23} = $& 1 &     0&  1&  0&  0& -1& -1\\
${\cal D}_4$&  0& $S_3      = $& 1 &     0&  0&  0&  1& -1& -1\\
${\cal D}_5$&  0& $S_1      = $& 1 &     0&  0&  1&  0& -1& -1\\
${\cal D}_6$&  0& $\bP_{12} = $& 1 &     0&  0&  0&  0& -1& -1\\
${\cal D}_7$&  0& $\Phi_1   = $& 1 &     0&  0&  0&  0&  0&  0\\
${\cal D}_8$&  0& $\Phi_2   = $& 1 &     0&  0&  0&  0&  0&  0\\
${\cal D}_9$&  0& $\Phi_3   = $& 1 &     0&  0&  0&  0&  0&  0\\
\hline
\end{tabular}
\end{flushleft}
\vspace{0.2cm}

In Tables 4.1 and 4.2, the first entry in a given row denotes the $D$-flat basis element label, 
the second entry denotes its anomalous charge (normalized to
${Q^{(A)}}'={Q^{(A)}}/16$),
and the remaining entries denote the ratios of the squares of the norms of its field VEVs.
The field corresponding to the first norm-square (that given in the third column) is unique to
the given flat direction and can be used to denote it.  
The VEVs for the first nine basis directions ($\cD_{10}$ through $\cD_{24}$) are formed solely from     
non-Abelian singlet fields (henceforth referred to simply as singlets), while
the VEVs in the remaining 15 basis directions ($\cD_{10}$ through $\cD_{24}$)  
contain several non-Abelian VEVs. Each of the 24 $D$-flat basis directions 
contains a unique field VEV not present in any of the other basis direction. 
Thus each flat directions can be identified by its associated VEV.
 
The VEVs forming $\cD_1$ through $\cD_6$ are of singlets from vector-like pairs. 
Thus there are also corresponding basis vectors, 
denoted as $\bar{\cD}_i$ for $i=1$, ..., 6, formed from vector-partner VEVs. 
Since the corresponding charges in $\cD_i$ and $\bar{\cD}_i$ are of opposite sign, we can 
express the $\bar{\cD}_i$ as $-{\cD}_i$ (effectively allowing the ${\cD}_i$ 
norm-squared components to be negative).  
The combination of $\cD_i$, $\bar{\cD}_1$, $i=1$, ..., 6, 
and the trivial uncharged moduli field directions
$\cD_{6+l} = <\Phi_l>$, for $l\in\{1,2,3\}$,  
form a complete set of singlet $D$-flat VEVs.

By definition, physical $D$-flat directions are not allowed negative norm-squares of VEVs
for non-vector-like fields, while they are allowed to have negative norm VEVs 
for vector-like components. Since vector pairs have opposite signed charges, a 
negative norm-squared implies that the vector partner field acquires the VEV, rather than the field.

\vspace{0.2cm}
\no Table 4.2: $D$-Flat Non-Abelian VEV Basis Elements.
\begin{flushleft}
\begin{tabular}{|l|r|lr|rrrrrrrr|}
\hline
Dir& ${Q^{(A)}}'$ & identifying field 
                  & & $K_4      $& $\bar{K}^{'}_3$&$S_8$&$\bar{\Phi}_{31}$&$S_2$&$S_4$&$S_5$&$S_6$\\
     &            & 
                  & & $\bar{K}_4$&                &     &                 &     &     &     &     \\
\hline
${\cal D}_{10}$& 2  &$F_{1}/\bar{F}_{1} = $ & 15  &  9 & 18 & -6 &-4 &  0 & 0 &-10 &-1\\  
${\cal D}_{11}$& 2  &$F_{2} =             $ & 15  & -6 &  3 & -6 &-4 &  0 & 0 &  5 &-1\\  
${\cal D}_{12}$& 4  &$F_{3} =             $ & 30  &-12 &  6 &-12 & 7 &-15 & 0 & -5 &-2\\  
${\cal D}_{13}$& 4  &$F_{4} =             $ & 30  &-12 &  6 &-12 & 7 &  0 &15 & -5 &-2\\  
${\cal D}_{14}$& 1  &$\bar{F}^{'}_2 =     $ &  5  &  2 & -1 &  2 &-2 &  0 & 0 &  0 & 2\\  
${\cal D}_{15}$& 2  &$\bar{F}^{'}_3 =     $ & 10  &  4 & -2 &  4 & 1 &  5 & 0 & -5 &-6\\  
${\cal D}_{16}$& 2  &$\bar{F}^{'}_4 =     $ & 10  &  4 & -2 &  4 & 1 &  0 &-5 &  5 & 4\\  
${\cal D}_{17}$& 2  &$K_1 =               $ &  6  &  0 &  6 &  0 &-1 &  0 & 3 & -1 & 2\\  
${\cal D}_{18}$& 2  &$K_2 =               $ &  6  &  0 &  6 &  0 &-1 & -3 & 0 & -1 & 2\\  
${\cal D}_{19}$& 1  &$K_3 =               $ &  3  &  0 &  3 &  0 &-2 &  0 & 0 &  1 & 1\\  
${\cal D}_{20}$& 1  &$\bar{K}^{'}_1 =     $ &  2  &  0 & -2 &  0 & 1 &  0 &-1 &  1 & 0\\  
${\cal D}_{21}$& 1  &$\bar{K}^{'}_2 =     $ &  2  &  0 & -2 &  0 & 1 &  1 & 0 & -1 &-2\\  
${\cal D}_{22}$& 0  &$N_1 =               $ &  2  & -2 & -2 &  2 &-1 &  0 & 1 &  1 & 0\\  
${\cal D}_{22}$& 0  &$N_2 =               $ &  2  & -2 & -2 &  2 &-1 & -1 & 0 &  3 & 2\\
${\cal D}_{24}$& 0  &$N_3 =               $ &  1  & -1 & -1 &  1 & 0 &  0 & 0 &  1 & 0\\
\hline
\end{tabular}
\end{flushleft}

Tables 4.1 and 4.2 reveal one property that all physical $D$-flat 
directions possess: ${\cal{D}}_1$ always appears (with positive coefficient), since only ${\cal{D}}_1$ carries a 
negative anomalous charge, necessary to cancel the positive FI-term. This is obvious when only singlet
flat directions are allowed since ${\cal{D}}_{2}$ to ${\cal{D}}_{9}$ carry no anomalous charge.
Thus the field $S_9$ acquires a VEV in all $D$-flat directions.

Inclusion of non-Abelian $D$-flat 
directions does not change this conclusion. For proof of this, 
first note that all 15 non-Abelian flat directions, i.e., 
${\cal{D}}_{10}$ to ${\cal{D}}_{24}$ carry positive anomalous charge.
Further, the fields unique 
to ${\cal{D}}_{11}$ to ${\cal{D}}_{24}$ are non-vector-like. 
Thus, $\bar{{\cal{D}}}_{11}$ to $\bar{{\cal{D}}}_{24}$ cannot appear in physical $D$ flat directions
with negative coefficients. Hence their $D$-term contributions are of the same sign as the FI-term.
Next, note that while ${\cal{D}}_{10}$'s unique field VEV, $F_1$, does have a vector partner,
$\bar{F}_1$), ${\cal{D}}_{10}$ also contains the VEV of the non-vector-like field $\bar{K}^{'}_3$.
In a physical direction the norm-square of the $\bar{K}^{'}_3$ VEV must be 
positive.
While the net norm-square the $\bar{K}^{'}_3$ VEV can be made positive by adding to ${\cal{D}}_{10}$ 
a linear combinations of basis directions 
${\cal{D}}_{11}$, ${\cal{D}}_{12}$, ${\cal{D}}_{13}$, ${\cal{D}}_{17}$, 
${\cal{D}}_{18}$, or ${\cal{D}}_{19}$, Table 4.2 shows that the net anomalous charge from 
such a combination turns positive again.
Hence a sufficiently large negative anomalous charge contribution from ${\cal{D}}_{1}$ is again required. 
Thus, ${\cal{D}}_{1}$ must be present in all valid $D$-flat directions, independent of 
non-Abelian field VEVs. In the following pages,
the phenomenological effect of a non-zero $<S_9>$ will often be discussed.
 
\subsection{Stringent $F$-flat Directions}

Linear combinations of the $D$-flat basis generators were systematically examined for stringent 
$F$-flatness in two steps: First the $D$-flat linear combinations were tested for 
stringent $F$-flatness through sixth order.\footnote{The number of superpotential terms increases drastically 
per order after sixth order in the superpotential, so this first test was used to limit the number of directions tested 
beyond sixth order.} (The optical unification model's superpotential is given up to sixth order in Appendix B.)
Then, directions passing the first test were  
examined for either all-order (or at least 17$^{\rm th}$) order stringent $F$-flatness. 
For those not $F$-flat to all (17$^{\rm th}$) order, 
the exact order at which $F$-flatness breaking occurs was determined.
Singlet flat directions were analyzed first, then flat directions containing non-Abelian VEVs. 

We searched among our stringent $F$-flat directions for those that   
induce FI-scale masses (see Appenedix C) to the MSSM-charged exotics that do not participate in the 
optical unification ``lensing'' effect. These unwanted MSSM exotics are 
(i) the six extra pairs of exotic $Q_Y=\pm \frac{1}{2}$ singlets and 
(ii) the three extra pairs of Higgs. 
For optical unification these states must acquire $\lambda_U$ or higher masses,
while, simultaneously, four exotic triplet/anti-triplets pairs, 
three exotic doublet pairs, and one singlet pairs
remains massless till an intermediate scale $\lambda_I$.

\subsubsection{Singlet Flat Directions}

Initially we allowed only non-Abelian singlet fields to take on VEVs. 
That is, we used only 
$\cD_i$, $i=1$, ..., 9, and $\bar{\cD}_j = -{\cD}_j $, $j= 2$, ..., 6,   
as our initial basis set.
The range of coefficients was from 0 to n for ${\cal D}_1$ and
from -n to n for ${\cal D}_{2\le i \le 6}$,
where n = 99 for directions containing up to four basis directions, 
n = 31 for directions containing five or six basis vectors,
and n = 21 for directions containing seven or more basis directions.
The coefficients for $\cD_{7,8,9}$ were either 0 or 1.
 
Note that $\cD_4$ has two unique field VEVs, $<S_1>$ and $<S_2>$.
Similarly, only $\cD_5$ contains $<S_3>$ and $<S_4>$. The superpotential
contains terms $S_1 S_2 \Phi_{12}$ and $S_3 S_4 \Phi_{12}$, 
which thus prohibit $\cD_4$ or $\cD_5$ 
from contributing to any stringently $F$-flat direction. 
$\bar{\cD}_4$ and $\bar{\cD}_5$ are similarly prohibited due to 
superpotential terms
$\bS_1 \bS_2 \bP_{12}$ and $\bS_3 \bS_4 \bP_{12}$.  
Linearly combinations of 
\beqn
n_1 \cD_1 + n_2 \cD_2 + n_3 \cD_3 + n_6 \cD_6    
\label{sing2}
\eeqn
are also constrained by the requirement that $< S_5 S_6> = < \bar{S}_5 \bar{S}_6> = 0$.
This requires that
\beqn
n_1 + n_2 + n_3 + n_6 = 0  
\label{sing3}
\eeqn
Ultimately, we found that the demand of stringent flatness through sixth order allowed 
only one class of singlet $D$-flat directions that is stringently $F$-flat to all order. 
All other $D$-flat directions generated where found to 
break $F$-flatness below seventh order in the superpotential.

The non-zero field VEVs for this all-order flat class are: 
\beqn
<S_9 \bS_7 \bS_8 > +\,\, {\rm one~or~more~of}\,\, 
<\bP_{12}>,\, <\bP_{31}>,\, <\P_2>,\, <\P_3>.
\label{sfset}
\eeqn
The specific set of all order flat directions is given in Table 4.3:

\vskip 0.2truecm
{\no Table 4.3: Norm-Squared Components of $D$-Flat Singlet VEVs with 
All-Order Stringent $F$-Flatness.
\begin{flushleft}
\begin{tabular}{|r|rrrrr|rr|}
\hline
${Q^{(A)}}'$ & $S_9$ & $\bS_7$ & $\bS_8$& $\bP_{12}$& $\bP_{31}$ & $\P_2$ & $\P_3$ \\
\hline
-1  & 3  &  2 &  1 &  1 & 0 & $v_2$ & $v_{3}$\\  
-1  & 3  &  1 &  2 &  0 & 1 & $v_2$ & $v_{3}$\\  
-2  & 6  &  3 &  3 &  1 & 1 & $v_2$ & $v_{3}$\\  
\hline
\end{tabular}
\end{flushleft}}
\no $v_{2}$ and $v_{3}$ are real, positive, and of FI scale, but otherwise unconstrained. 

\subsubsection{Singlet Flat Direction Class Phenomenology}

For the singlet class of flat directions defined by (\ref{sfset}),
no (rather than a hoped for six) linear combinations 
of the seven pairs of singlets with $Q_Y=\pm \frac{1}{2}$ receive mass (see Table C.3).
From Table C.1a we also find that the VEVs of $\P_2$ and $\P_3$ should be kept at zero 
to prevent third and fifth order mass terms
\beqn
\half <\P_3> D_1 \bD_1,\, 
\half <\P_3> D_2 \bD_2,\,
\half <\P_2 > D_4 \bD_4,\,  
<S_9 \bar{S}_8 \P_2 > D_1 \bD_2.
\label{d44ml} 
\eeqn
from appearing for the exotic $D$ triplets.
The mass terms in (\ref{d44ml}) should be zero, 
for it is extremely unlikely that the third order terms could be cancelled
by the possible sixth order terms in respective
$m^{'}_{11}$, $m^{'}_{22}$, and $m^{'}_{44}$, since 
sixth order terms have order(1/100) suppression. 
Further, there is a similar suppression ratio between the fifth order term in (\ref{d44ml})
and possible seventh order contributions. 

Third order stringent $F$-flatness constraints also forbid both fields of a given vector pair 
from simultaneously acquiring VEVs. That is,
\beqn
<F_1 \bF_1> = <K_4 \bK_4> = <S_i \bS_i> = 0.
\label{vpv}   
\eeqn
Thus, (\ref{sfset}) implies that
\beqn
<S_7> = <S_8> = <\bS_9> = 0.  
\label{novset1}
\eeqn
(Third order constraints also require that
\beqn
<S_7 \bS_8> = <S_8 \bS_7> = 0,
\label{vpv2}   
\eeqn
but (\ref{vpv2}) is automatically satisfied when (\ref{vpv}) is combined with (\ref{sfset}).)
Note also that (\ref{novset1}) 
removes the third and higher order $D\bD$ mass terms containing $S_8$ or $\bS_9$.
Hence, for stringent flat directions,
all $D_2\bD_1$ terms vanish (when combined with $<\P_1> = <\P_2> = 0$), 
as do third order and several fifth order $D_2\bD_4$ terms, 
and the sixth order $D_3\bD_3$ terms. Therefore, 
the only $D\bD$ that need be investigated for non-Abelian stringent directions
are those in $m^{'}_{11}$, $m^{'}_{22}$, the last half of $m^{'}_{24}$ and $m^{'}_{41}$ terms, and 
$m^{'}_{44}$. 
 
The singlet class flat directions do not give unwanted mass to
the exotic $X_i$ and $\bar{X}_i$ doublets through at least sixth order. 
Further, by demanding   
\beqn
<\P_2> = 0
\label{p2z}
\eeqn
because of (\ref{d44ml}), 
the unwanted possible fifth order $X_1 \bX_1$ mass term from Table C.2 is also 
eliminated, independent of  
$<S_1 \bS_6 + S_1 \bS_6 >$, which may not be zero for generic non-Abelian directions. 
Unfortunately, Table C.3a also reveals that (\ref{p2z}) eliminates the desirable mass terms 
(of fifth order) for $A_{4} \bA{4}$ and $A_{7} \bA{7}$ and for $A_{7} \bA{1}$ 
(of sixth order). 

Note that $\P_1$ was not allowed a VEV in any singlet flat direction because of the 
third order term, $S_9 \bS_9 \P_{1}$ (and also because of $S_7 \bS_7 \P_{1}$ and $S_8 \bS_8 \P_{1}$).
Since $<S_9> \ne 0$ also applies to all non-Abelian flat directions, $<\P_1> = 0$ is also true
for all flat directions. 
$<\P_1> = 0$ (favorably) prevents a possible $X_1 \bX_1$ mass term from appearing for the 
latter directions. 
However, as Table C.3a indicates, this also  prevents the desirable 
fifth order mass terms for exotic singlets $A_2 \bA_2$ and $A_5 \bA_5$, 
which means that, at most, one independent $A \bA$ mass term can be expected from sixth order or lower
non-Abelian flat directions,
\beqn
(<F_1 \bF^{'}_2 N_3 S_9> A_1 + <N_2> A_5)\bA_1.
\label{aam1}
\eeqn
One possible difficulty with this is both mass components require left-handed anti-neutrino 
singlet VEVs, which might result in unacceptable lepton number violation.

This analysis implies that successful optical unification clearly requires seventh order mass terms for
$A \bA$ and possibly for some $A A$ (see Table C.3b).
Since these masses must be at or above the $\lambda_U$ scale, the WCHS unification scale $\lambda_H$ for this model
must be above the lower bound of $5\times 10^{17}$ GeV and on the order of $10^{18}$ GeV $\times$ Order(1), 
In addition, several seventh order mass terms may also be required   
for a given $A_i \bA_j$ or $A_i A_j$, in order to sufficiently counter seventh order mass suppression. 
(Whether this occurs or not will be studied in the next section.)

We complete our discussion of singlet direction phenomenology with an analysis of MSSM mass matrices:
From the Higgs mass matrix given in Table C.4a, we find that $<S_9>$ produces the term
\beqn 
<S_9 > h_1 \bh_4.
\label{gh14}
\eeqn 
Additionally, the
non-zero optional singlet VEVs would yield a second (linearly independent) combination  
\beqn
h_1 ( <\bP_{12}> \bh_2 + <\bP_{31}> \bh_3 ), 
\label{gh12}
\eeqn 
leaving but one more desirable Higgs mass term to be generated.

Third order diagonal up quark mass terms of the form, 
\beqn
Q_i u^c_i \bh_i,
\label{hmugi}
\eeqn
generically appear in NAHE-based models \cite{af3}.
These can naturally produce a generational mass hierarchy if
each $\bh_i$ appears in the physical massless Higgs combination with coefficients
differing by orders of magnitude. Mass hierarchy between two generations is often 
obtained this way in NAHE-based models, but generally not for all three generation. 
Rather, additional suppressed quark mass terms are needed.
Depending on the optional VEVs acquired,
the singlet VEV class can provide suppressed mass terms 
\beqn
Q_2 u^c_2 \bh_4 < S_9 \bP_{12}>,\,\, {\rm and}\,\, Q_3 u^c_3 \bh_4 < S_9 \bP_{31}>
\label{umsi}
\eeqn 
(see Table C.5).
Note that the three suppressed sixth order $Q_1 d^c_1 h_4$ mass terms and a similar set 
for $Q_2 d^c_2 h_4$ are prohibited by stringent third order $F$-flatness.  

Natural mass suppression between up and down quarks becomes evident with Table C.6, which 
does not contain comparable
$Q_i d^c_i h_i$ terms, which 
follows the pattern discussed in \cite{nmssm,af3}. Based on GSO projection
choices, third order mass terms for either up quarks or down quarks (but not both) can appear. 
The singlet flat direction class does provide one fourth order down quark mass term,
\beqn
Q_2 d^c_2 h_4 < \bS_8>,
\label{dmus22}
\eeqn
and possibly one suppressed sixth order term,
\beqn
Q_2 d^c_2 h_3 < S_9 \bS_7 \bP_{12}>.
\label{umus22}
\eeqn
As with the up quarks, the three suppressed sixth order $Q_1 d^c_1 h_4$ mass terms 
are prohibited by stringent third order $F$-flatness.  Further, for upcoming non-Abelian 
directions, the fifth order $Q_2 d^c_2 h_2$ terms are similarly eliminated. 
Lastly, note that Table C.6 is also the electron mass matrix, providing
$m_{e_i} = m_{d_i}$ at the string scale. 
  
As already discussed, other than the singlet $D$-flat direction class (\ref{sfset}), 
all singlet $D$-flat directions lose $F$-flatness below sixth order. 
Thus, the insufficiency of the (\ref{sfset}) singlet flat-direction class, i.e., its lack of 
mass terms for the six extra pairs of singlets, led us to investigate
the phenomenologies of non-Abelian stringent flat directions, containing VEVs of hidden-sector   
$5$ and $\bar{5}$ fields of $SU(5)_H$ and/or the $3$ and $\bar{3}$ fields of $SU(3)_H$.  

\subsubsection{Non-Abelian Flat Directions}

We systematically generated non-Abelian flat directions using the basis directions from both Tables 4.1 and 4.2,
with at least one direction always from Table 4.2.
Under reasonable constraints for range of basis vector coefficients
with regard to program running time,
the complete collection of non-Abelian stringent flat directions for our model
was generated and analyzed.  Linear combinations of up to seven of the basis directions were examined. 
For linear combinations of three or four basis directions, an integer range of coefficients 
from -99 to 99 was chosen, 
whereas for five or six basis directions, a reduced coefficient range from -31 to 31 was used, 
and for seven basis directions, a further reduced coefficient range from -21 to 21 was applied.
The maximum number of basis vector considered so far was seven because of two factors: the projected running time
for eight basis vectors is several weeks and no new stringently flat direction classes 
(or self-cancellation possibilities) were found for seven basis vectors. 

Our investigation revealed
five classes of all-order (or at least 17$^{\rm th}$ order) stringently flat directions
and one class that can be made stringently flat by self-cancellation of non-Abelian field VEVs 
(see Table D.1). Classes are denoted by a $n-d$ label, where $n$ is the number of independent pairs of $A_i \bA_j$ 
exotic doublets that gain mass and $d$ 
designates different combinations of the $n$ pairs. We found three classes that  
provide only one independent MSSM exotic singlet mass term, one class that provides two independent mass terms,
and two classes that provide three independent mass terms.

The thirty all-order stringent flat directions in class 1-1 generate third order mass for $A_5\bA_1$
and contain anywhere from 8 to 11 field VEVs. 
The four all-order stringent flat directions in class 1-2 class generate seventh order mass  for $A_7 \bA_6$ 
and contain either ten or eleven VEVs. 
The single all-order stringent flat direction forming class 1-3 yields seventh order mass for $A_3 \bA_4$. 
The four stringent flat directions in class 2-1 produce a third order mass term for 
$A_5 \bA_1$ and seventh order mass terms for $A_5 \bA_7$ and $A_2 \bA_4$, which generate
mass for $A_5$ and a linear combination of $\lambda_3 \bA_1 + \lambda_7 \bA_7$. 
The single all-order stringent flat direction in class 3-1 contributes seventh order masses to 
$A_4 \bA_2$, $A_7 \bA_5$, and $A_6 \bA_1$. 

Half of the class 2-1 directions have phenomenological difficulties:
flat direction 2-1.2 generates an unwanted seventh order mass for the exotic doublets  
$\bX_1 \bX_1$, while 2-1.4 generates an unwanted sixth order mass term for the exotic triplets $D_4 \bD_1$. 
Additionally, 3-1.1 produces unwanted 7th order mass for $\bX_2 \bX_2$.
Thus, although stringent flat direction 3-1.1 can produce MSSM scale or higher mass for 
three pairs of $A$/$\bA$ fields, the better phenomenological starting point for
a flat direction is either flat direction 2-1.1 or 2-1.3, both of which keep all of 
the exotic $D/\bD$ triplets and the exotic $X/\bX$ doublets massless at the MSSM-scale.  

Self-cancellation of the class 3-2 directions in Table D.3, providing  
stringent flatness to at least 17$^{\rm th}$ order, results in a further improved starting point
for optical unification.
The four directions in class 3-2 provide mass for three exotic doublet pairs, 
$(a_3 A_5 + a_7 A_6)\bA_1$, $A_2\bA_4$, and $A_5\bA_7$, 
while simultaneously keeping all $D$/$\bD$ and $X$/$\bX$ fields massless.
Note that the mass terms in classes 1-1 and 2-1 are subsets of the class 3-2 set, 
while the addition of class 1-3 to 3-2 simply rotates an
$A$ mass eigenstate. Linear combinations of 1-2 and 3-2, requiring at least 8 basis vectors, 
generate four independent exotic singlet mass terms (and are discussed further below).

All four flat directions in class 3-2 are threatened by varying numbers of 16$^{\rm th}$ order superpotential 
terms and several derived $F$-terms.\footnote{16$^{\rm th}$ order is still 
likely unacceptably low by one order, producing SUSY breaking at an energy scale too high 
by approximately a factor of ten.} 
Nevertheless, the dangerous $<W>$ and $<F>$ terms can be eliminated for all four directions by 
non-Abelian self-cancellation.
For example, flat direction 3-2.1 (representative of this class) is endangered by $W$-term
\beqn
&& {S_9}^3 {\bS_8}^2 \bP_{31} {\bS_2}^2 S_6 N_2 (K_2\cdot\bK^{'}_3)^2 (K_2\cdot\bK_4) \label{fb1}
\eeqn
and related $F$-terms.
In flat direction 3-2.1, the squares of the norms of the non-Abelian VEVs are
\beqn
<K_1>^2 = <K_2>^2 = <\bK^{'}_1>^2 = 3 <\bK^{'}_3>^2 = 3/2 <\bK_4 >^2 = 6.
\label{fd361}
\eeqn 
One VEV choice for maintaining hidden $SU(3)$ $D$-term flatness is
\beqn
&& <K_1> = < \sqrt{6},\, 0,\, 0 >,\, 
<\bK^{'}_3> = < \sqrt{2},\, 0,\, 0 >,\, 
<\bK_4 > = < 2,\, 0,\, 0 >,
\nolabel\\ 
&& <K_2> =  < \sqrt{6},\, 0,\, 0 >,\quad
<\bK^{'}_1> < \sqrt{6},\, 0,\, 0 >.
\label{fd361b}
\eeqn 
This provides for,
\beqn
(K_2 \cdot \bK^{'}_{3}) = 0,
\label{cancelkbk}
\eeqn
which eliminates all dangerous $W$- and $F$-terms.  

As with the singlet flat directions, all of the above six classes of non-Abelian directions also generate:
(i) third order mass terms for two of the four Higgs doublets, $<S_9> h_1 \bh_4$ and $<\bP_{31}> h_3 \bh_1$,
(ii) a suppressed fifth order up quark mass term, $<S_9 \bP_{31}> \bh_{4} u_3 \bar{u}_3$, and 
(iii) matching fourth order down quark and electron mass terms, $<\bS_8> h_{4} (d_2 \bar{d}_2 + e^{-}_2 e^{+}_2)$.
 
As for the linear combinations of classes 1-2 and 3-2, discussed prior (which generate four independent 
mass terms): all were found to lose flatness by two fourteenth order terms, 
\beqn
{S_9}^3 {\bS_8}^2 \bP_{31} \bS_6 N_2 [(K_2\cdot {\bK^{'}_2})(K_3\cdot\bK^{'}_2) (K_3\cdot\bK_4) + 
                                      (K_2\cdot {\bK^{'}_4})(K_3\cdot\bK^{'}_2)^2 ].
\label{w14bk8}
\eeqn
Self-cancellation of fifth, sixth, and tenth order terms endangering $F$-flatness of these linear combinations 
require
\beqn 
<K_3\cdot {\bK^{'}_3} > = <K_1\cdot \bK_4 > = 0, 
\label{w5610bk}
\eeqn
which can be shown to be consistent only with self-cancellation of the first term in (\ref{w14bk8}).

Several directions that generate mass for not just three, but five independent sets of 
$A\bA$ were also found (see Table D.4). Four representatives of class 5-1 and four of class 5-2 are given. 
Unfortunately, all class 5-1 and 5-2 directions are broken by an eleventh order superpotential term,
\beqn
{S_9}^2 {\bS_8}^2 \bP_{31} \bS_2 S_5 (K_2\cdot \bK^{'}_3)^2.
\label{w11bk}
\eeqn
None of these directions contain two pair of $K$ and $\bK$, and thus the self-cancellation condition of
(\ref{cancelkbk}) cannot be imposed.

\subsection{General Flat Direction Investigation} 

Several stringent flat directions, 
including some kept $F$-flat by self-cancellation of sixteenth order
superpotential terms, were found that generate $\Lambda_U$ scale or higher mass 
for up to half of the extra pairs of hypercharge-carrying exotic singlets (the $A$ and $\bA$) that 
cannot contribute to optical unification. 
Many of these stringent flat directions keep all four pairs of the exotic triplets (the $D$ and $\bD$) and 
all three pairs of the exotic doublets (the $X$ and $\bX)$ massless at or below $\Lambda_U$. 
However, no directions stringently flat to at least 17$^{\rm th}$ order have yet been found that 
can provide $\Lambda_U$ scale mass to four, five, or all six of the extra exotic singlets. 
Stringent directions generating five mass pairs were found, but these lose $F$-flatness at no higher than 11$^{\rm th}$ order.

A systematic search for stringent flat directions that give mass to more than three exotic singlet pairs is underway 
This search involves using eight or more $D$-flat basis directions and run time will be several weeks. 
That stringent flat directions giving mass to four or more of the exotic singlet pairs have not been found
suggests that if the model under investigation is to realize optical unification, 
a non-stringent flat direction is likely required. $F$-flatness requirements suggest that
such a direction will likely have a stringent flat direction embedded within as a root.
Thus, future research will focus on a systematic search for non-stringent $F$-flat variations derived from stringent
directions. 
 
\section{Summary}

In the context of the weakly coupled heterotic string (WCHS)
and the likelihood of intermediate scale MSSM-charged exotics (ISME) in realistic models,
optical unification offers an explanation for the perhaps otherwise apparently accidental  
unification of the MSSM running couplings at
$\Lambda_U \sim 2.5 \times 10^{16}$ GeV, for the intermediate scale desert scenario (ISD),
rather than at or above the lower limit of the 
string coupling unification scale, $\Lambda_H \sim 5 \times 10^{17}$ GeV.
For a set of ISME particles meeting optical unification constraints,
a virtual MSSM unification scale below the real string unification scale is guaranteed.

A WCHS model of the NAHE class has been found that offers 
possible realization of optical unification. In this model optical unification requires that
six pairs of exotic non-Abelian singlet states with zero hypercharge acquire
$\Lambda_U$ scale  or larger masses, along with three out of four pairs of MSSM Higgs. Additionally,
the four pairs of exotic MSSM triples and three pairs of MSSM doublets must remain massless down to 
an intermeidate scale $\Lambda_I$.
The optical unification properties of systematically generated 
$D$- and stringent $F$-flat directions have been investigated for this model.
As possible roots for more general $F$-flat directions, the stringent directions were found to 
provide $\Lambda_U$ scale mass 
for up to three of the six unwanted exotic singlets and two of the four pairs of MSSM Higgs.
Investigations of non-stringent $F$-flat directions that might generate all six desired 
exotic singlet masses have been initiated. 

\section*{Acknowledgments}
This paper is a result of the 2003 and 2004 NSF
REU summer programs at Baylor University. 
Funding for the REU students was provided by NSF grant no.~0097386. 
Research for this paper was supported in part by the NASA/Texas Space Grant Consortium.
\vfill
\newpage

\def\AEF{A.~E.~Faraggi}
\def\AP#1#2#3{{\it Ann.\ Phys.}\/ {\bf#1} (#2) #3}
\def\NPB#1#2#3{{\it Nucl.\ Phys.}\/ {\bf B#1} (#2) #3}
\def\NPBPS#1#2#3{{\it Nucl.\ Phys.}\/ {{\bf B} (Proc. Suppl.) {\bf #1}} (#2)
 #3}
\def\PLB#1#2#3{{\it Phys.\ Lett.}\/ {\bf B#1} (#2) #3}
\def\PRD#1#2#3{{\it Phys.\ Rev.}\/ {\bf D#1} (#2) #3}
\def\PRL#1#2#3{{\it Phys.\ Rev.\ Lett.}\/ {\bf #1} (#2) #3}
\def\PRT#1#2#3{{\it Phys.\ Rep.}\/ {\bf#1} (#2) #3}
\def\PTP#1#2#3{{\it Prog.\ Theo.\ Phys.}\/ {\bf#1} (#2) #3}
\def\MODA#1#2#3{{\it Mod.\ Phys.\ Lett.}\/ {\bf A#1} (#2) #3}
\def\MPLA#1#2#3{{\it Mod.\ Phys.\ Lett.}\/ {\bf A#1} (#2) #3}
\def\IJMP#1#2#3{{\it Int.\ J.\ Mod.\ Phys.}\/ {\bf A#1} (#2) #3}
\def\IJMPA#1#2#3{{\it Int.\ J.\ Mod.\ Phys.}\/ {\bf A#1} (#2) #3}
\def\JHEP#1#2#3{{\it JHEP}\/ {\bf #1} (#2) #3}
\def\nuvc#1#2#3{{\it Nuovo Cimento}\/ {\bf #1A} (#2) #3}
\def\RPP#1#2#3{{\it Rept.\ Prog.\ Phys.}\/ {\bf #1} (#2) #3}
\def\etal{{\it et al\/}}


\newpage

\appendix

\section{Optical Unification Model Fields and Their Charges}

\begin{flushleft}
\noindent Table A.1: MSSM 3 Generations \& Higgs
\begin{tabular}{|l||c|cccccccc|c|cc|}
\hline
  $F$      & $(SU(3)_C,$ & $Q_{Y}$ & $Q_{Z'}$
   & $Q_A$ & $Q_{1}$ & $Q_{2}$ & $Q_{3}$         & $Q_{4}$
           & $Q_{5}$ & $(SU(5)_{H},$   & $Q_{6}$ & $Q_{7}$  \\
           & $SU(2)_L)$ & & & & & & & & & $SU(3)_{H})$ & &  \\
\hline
      $Q_1$& $(3,2)$        & 1/6 & 1/6 & 1/2 &-1/2 &1/2 & 0   & 1/2 & 0   & $(1,1)$ & 0 & 0\\
      $u_1$& $(1,2)$        &-2/3 & 1/3 & 1/2 &-1/2 &1/2 & 0   &-1/2 & 0   & $(1,1)$ & 0 & 0\\
      $d_1$& $(1,2)$        & 1/3 &-2/3 & 1/2 &-1/2 &1/2 & 0   &-1/2 & 0   & $(1,1)$ & 0 & 0\\
      $L_1$& $(1,2)$        &-1/2 &-1/2 & 1/2 &-1/2 &1/2 & 0   & 1/2 & 0   & $(1,1)$ & 0 & 0\\
      $e_1$& $(1,2)$        & 1   & 0   & 1/2 &-1/2 &1/2 & 0   &-1/2 & 0   & $(1,1)$ & 0 & 0\\
      $N_1$& $(1,2)$        & 0   & 1   & 1/2 &-1/2 &1/2 & 0   &-1/2 & 0   & $(1,1)$ & 0 & 0\\
\hline
      $Q_2$& $(3,2)$        & 1/6 & 1/6 & 1/2 & 1/2 &1/2 &-1/2 & 0   & 0   & $(1,1)$ & 0 & 0\\
      $u_2$& $(1,2)$        &-2/3 & 1/3 & 1/2 & 1/2 &1/2 & 1/2 & 0   & 0   & $(1,1)$ & 0 & 0\\
      $d_2$& $(1,2)$        & 1/3 &-2/3 & 1/2 & 1/2 &1/2 & 1/2 & 0   & 0   & $(1,1)$ & 0 & 0\\
      $L_2$& $(1,2)$        &-1/2 &-1/2 & 1/2 & 1/2 &1/2 &-1/2 & 0   & 0   & $(1,1)$ & 0 & 0\\
      $e_2$& $(1,2)$        & 1   & 0   & 1/2 & 1/2 &1/2 & 1/2 & 0   & 0   & $(1,1)$ & 0 & 0\\
      $N_2$& $(1,2)$        & 0   & 1   & 1/2 & 1/2 &1/2 & 1/2 & 0   & 0   & $(1,1)$ & 0 & 0\\
\hline
      $Q_3$& $(3,2)$        & 1/6 & 1/6 & 1/2 & 0   &-1  & 0   & 0   &-1/2 & $(1,1)$ & 0 & 0\\
      $u_3$& $(1,2)$        &-2/3 & 1/3 & 1/2 & 0   &-1  & 0   & 0   & 1/2 & $(1,1)$ & 0 & 0\\
      $d_3$& $(1,2)$        & 1/3 &-2/3 & 1/2 & 0   &-1  & 0   & 0   & 1/2 & $(1,1)$ & 0 & 0\\
      $L_3$& $(1,2)$        &-1/2 &-1/2 & 1/2 & 0   &-1  & 0   & 0   &-1/2 & $(1,1)$ & 0 & 0\\
      $e_3$& $(1,2)$        & 1   & 0   & 1/2 & 0   &-1  & 0   & 0   & 1/2 & $(1,1)$ & 0 & 0\\
      $N_3$& $(1,2)$        & 0   & 1   & 1/2 & 0   &-1  & 0   & 0   & 1/2 & $(1,1)$ & 0 & 0\\
\hline
      $h_1$& $(1,2)$        &-1/2 & 1/2 & 1  &-1  & 1  & 0 & 0 & 0 & $(1,1)$ & 0 & 0\\
      $h_2$& $(1,2)$        &-1/2 & 1/2 & 1  & 1  & 1  & 0 & 0 & 0 & $(1,1)$ & 0 & 0\\
      $h_3$& $(1,2)$        &-1/2 & 1/2 & 1  & 0  &-2  & 0 & 0 & 0 & $(1,1)$ & 0 & 0\\
      $h_4$& $(1,2)$        &-1/2 & 0   &-1/4&-1/2&1/2 & 0 & 0 & 0 & $(1,1)$ & 2 & 0\\
$\bar{h}_1$& $(1,2)$        & 1/2 &-1/2 &-1  & 1  &-1  & 0 & 0 & 0 & $(1,1)$ & 0 & 0\\
$\bar{h}_2$& $(1,2)$        & 1/2 &-1/2 &-1  &-1  &-1  & 0 & 0 & 0 & $(1,1)$ & 0 & 0\\
$\bar{h}_3$& $(1,2)$        & 1/2 &-1/2 &-1  & 0  & 2  & 0 & 0 & 0 & $(1,1)$ & 0 & 0\\
$\bar{h}_4$& $(1,2)$        & 1/2 & 0   &1/4 &1/2 &-1/2& 0 & 0 & 0 & $(1,1)$ &-2 & 0\\
\hline
\end{tabular}
\vfill
\newpage

\no Table A.2: MSSM-Charged Exotics
\begin{tabular}{|l||c|cccccccc|c|cc|}
\hline
  $F$      & $(SU(3)_C,$ & $Q_{Y}$ & $Q_{Z'}$
   & $Q_A$ & $Q_{1}$ & $Q_{2}$ & $Q_{3}$         & $Q_{4}$
           & $Q_{5}$ & $(SU(5)_{H},$   & $Q_{6}$ & $Q_{7}$  \\
           & $SU(2)_L)$ & & & & & & & & & $SU(3)_{H})$ & &  \\
\hline
      $D_1$& $(3,1)$        &-1/3 &-1/3 & 1  & 0  & 1  & 0  &  0 & 0  & $(1,1)$ & 0 & 0 \\
      $D_2$& $(3,1)$        &-1/3 &-1/3 &-1  & 0  &-1  & 0  &  0 & 0  & $(1,1)$ & 0 & 0 \\
      $D_3$& $(3,1)$        &-1/3 & 1/6 & 1/4&-1/2&-1/2& 0  &  0 & 0  & $(1,1)$ &-2 & 0 \\
      $D_4$& $(3,1)$        & 1/6 & 1/6 & 0  & 0  & 0  &1/2 & 1/2& 1/2& $(1,1)$ &1/2&-15/2\\
$\bar{D}_1$& $({\bar 3},1)$ & 1/3 & 1/3 &-1  & 0  &-1  & 0  &  0 & 0  & $(1,1)$ & 0 & 0 \\
$\bar{D}_2$& $({\bar 3},1)$ & 1/3 & 1/3 & 1  & 0  & 1  & 0  &  0 & 0  & $(1,1)$ & 0 & 0 \\
$\bar{D}_3$& $({\bar 3},1)$ & 1/3 & 1/6 &-1/4&1/2 &1/2 & 0  &  0 & 0  & $(1,1)$ & 2 & 0 \\
$\bar{D}_4$& $({\bar 3},1)$ &-1/6 &-1/6 & 0  & 0  & 0  &-1/2&-1/2&-1/2& $(1,1)$ &-1/2& 15/2\\
\hline
      $X_1$& $(1,2)$& 0   & 0   & 1/2&-1/2& 1/2&1/2& 0  &1/2& $(1,1)$ &-1/2& 15/2\\
      $X_2$& $(1,2)$& 0   & 0   & 1/2& 1/2& 1/2& 0 &-1/2&1/2& $(1,1)$ &-1/2& 15/2\\
      $X_3$& $(1,2)$& 0   & 0   & 1/2& 0  &-1  &1/2&-1/2& 0 & $(1,1)$ &-1/2& 15/2\\
$\bar{X}_1$& $(1,2)$& 0   & 0   &-1/2& 1/2&-1/2&1/2& 0  &1/2& $(1,1)$ & 1/2&-15/2\\
$\bar{X}_2$& $(1,2)$& 0   & 0   &-1/2&-1/2&-1/2& 0 &-1/2&1/2& $(1,1)$ & 1/2&-15/2\\
$\bar{X}_3$& $(1,2)$& 0   & 0   &-1/2& 0  & 1  &1/2&-1/2& 0 & $(1,1)$ & 1/2&-15/2\\
\hline
      $A_1$& $(1,1)$        & 1/2 & 1/2 & 0  & 0  & 0  & 1/2& 1/2&-1/2& $(1,1)$&-1/2& 15/2\\
      $A_2$& $(1,1)$        &-1/2 & 1/2 &-1/2&-1/2&-1/2& 0  & 1/2&-1/2& $(1,1)$&-1/2& 15/2\\
      $A_3$& $(1,1)$        &-1/2 & 1/2 &-1/2& 0  & 1  &-1/2& 1/2& 0  & $(1,1)$&-1/2& 15/2\\
      $A_4$& $(1,1)$        &-1/2 & 1/2 &-1/2& 1/2&-1/2&-1/2& 0  &-1/2& $(1,1)$&-1/2& 15/2\\
      $A_5$& $(1,1)$        & 1/2 &-1/2 &-1/2&-1/2&-1/2& 0  & 1/2&-1/2& $(1,1)$&-1/2& 15/2\\
      $A_6$& $(1,1)$        & 1/2 &-1/2 &-1/2& 0  & 1  &-1/2& 1/2& 0  & $(1,1)$&-1/2& 15/2\\
      $A_7$& $(1,1)$        & 1/2 &-1/2 &-1/2& 1/2&-1/2&-1/2& 0  &-1/2& $(1,1)$&-1/2& 15/2\\
$\bar{A}_1$& $(1,1)$        &-1/2 &-1/2 & 0  & 0  & 0  &-1/2&-1/2& 1/2& $(1,1)$& 1/2&-15/2\\
$\bar{A}_2$& $(1,1)$        & 1/2 &-1/2 & 1/2& 1/2& 1/2& 0  & 1/2&-1/2& $(1,1)$& 1/2&-15/2\\
$\bar{A}_3$& $(1,1)$        & 1/2 &-1/2 & 1/2& 0  &-1  &-1/2& 1/2& 0  & $(1,1)$& 1/2&-15/2\\
$\bar{A}_4$& $(1,1)$        & 1/2 &-1/2 & 1/2&-1/2& 1/2&-1/2& 0  &-1/2& $(1,1)$& 1/2&-15/2\\
$\bar{A}_5$& $(1,1)$        &-1/2 & 1/2 & 1/2& 1/2& 1/2& 0  & 1/2&-1/2& $(1,1)$& 1/2&-15/2\\
$\bar{A}_6$& $(1,1)$        &-1/2 & 1/2 & 1/2& 0  &-1  &-1/2& 1/2& 0  & $(1,1)$& 1/2&-15/2\\
$\bar{A}_7$& $(1,1)$        &-1/2 & 1/2 & 1/2&-1/2& 1/2&-1/2& 0  &-1/2& $(1,1)$& 1/2&-15/2\\
\hline
\end{tabular}
\vfill
\newpage

\no Table A.3: Singlets with $Q_Y=0$
\begin{tabular}{|l||c|cccccccc|c|cc|}
\hline
  $F$      & $(SU(3)_C,$ & $Q_{Y}$ & $Q_{Z'}$
   & $Q_A$ & $Q_{1}$ & $Q_{2}$ & $Q_{3}$         & $Q_{4}$
           & $Q_{5}$ & $(SU(5)_{H},$   & $Q_{6}$ & $Q_{7}$  \\
           & $SU(2)_L)$ & & & & & & & & & $SU(3)_{H})$ & &  \\
\hline
  $\Phi_1$ & $(1,1)$        & 0   & 0   & 0 & 0 & 0 & 0 & 0 & 0 & $(1,1)$ & 0 & 0\\
  $\Phi_2$ & $(1,1)$        & 0   & 0   & 0 & 0 & 0 & 0 & 0 & 0 & $(1,1)$ & 0 & 0\\
  $\Phi_3$ & $(1,1)$        & 0   & 0   & 0 & 0 & 0 & 0 & 0 & 0 & $(1,1)$ & 0 & 0\\
$\Phi_{12}$& $(1,1)$        & 0   & 0   & 0 &-2 & 0 & 0 & 0 & 0 & $(1,1)$ & 0 & 0\\
$\Phi_{23}$& $(1,1)$        & 0   & 0   & 0 & 1 &-3 & 0 & 0 & 0 & $(1,1)$ & 0 & 0\\
$\Phi_{31}$& $(1,1)$        & 0   & 0   & 0 &-1 &-3 & 0 & 0 & 0 & $(1,1)$ & 0 & 0\\
${\bar\Phi}_{12}$&$(1,1)$   & 0   & 0   & 0 & 2 & 0 & 0 & 0 & 0 & $(1,1)$ & 0 & 0\\
${\bar\Phi}_{23}$&$(1,1)$   & 0   & 0   & 0 &-1 & 3 & 0 & 0 & 0 & $(1,1)$ & 0 & 0\\
${\bar\Phi}_{31}$&$(1,1)$   & 0   & 0   & 0 & 1 & 3 & 0 & 0 & 0 & $(1,1)$ & 0 & 0\\
\hline
      $S_1$& $(1,1)$        & 0   & 0   & 0 &-1 & 0 &-1 & 0 & 0 & $(1,1)$ & 0 & 0\\
      $S_2$& $(1,1)$        & 0   & 0   & 0 &-1 & 0 & 1 & 0 & 0 & $(1,1)$ & 0 & 0\\
      $S_3$& $(1,1)$        & 0   & 0   & 0 &-1 & 0 & 0 &-1 & 0 & $(1,1)$ & 0 & 0\\
      $S_4$& $(1,1)$        & 0   & 0   & 0 &-1 & 0 & 0 & 1 & 0 & $(1,1)$ & 0 & 0\\
      $S_5$& $(1,1)$        & 0   & 0   & 0 &-1 & 0 & 0 & 0 &-1 & $(1,1)$ & 0 & 0\\
      $S_6$& $(1,1)$        & 0   & 0   & 0 &-1 & 0 & 0 & 0 & 1 & $(1,1)$ & 0 & 0\\
      $S_7$& $(1,1)$        & 0   & 1/2 &3/4&-1/2&-3/2&0& 0 & 0 & $(1,1)$ & 2 & 0\\
      $S_8$& $(1,1)$        & 0   & 1/2 &3/4&1/2&3/2& 0 & 0 & 0 & $(1,1)$ & 2 & 0\\
      $S_9$& $(1,1)$        & 0   & 1/2 &-5/4&1/2&-1/2&0& 0 & 0 & $(1,1)$ & 2 & 0\\
$\bar{S}_1$& $(1,1)$        & 0   & 0   & 0 & 1 & 0 & 1 & 0 & 0 & $(1,1)$ & 0 & 0\\
$\bar{S}_2$& $(1,1)$        & 0   & 0   & 0 & 1 & 0 &-1 & 0 & 0 & $(1,1)$ & 0 & 0\\
$\bar{S}_3$& $(1,1)$        & 0   & 0   & 0 & 1 & 0 & 0 & 1 & 0 & $(1,1)$ & 0 & 0\\
$\bar{S}_4$& $(1,1)$        & 0   & 0   & 0 & 1 & 0 & 0 &-1 & 0 & $(1,1)$ & 0 & 0\\
$\bar{S}_5$& $(1,1)$        & 0   & 0   & 0 & 1 & 0 & 0 & 0 & 1 & $(1,1)$ & 0 & 0\\
$\bar{S}_6$& $(1,1)$        & 0   & 0   & 0 & 1 & 0 & 0 & 0 &-1 & $(1,1)$ & 0 & 0\\
$\bar{S}_7$& $(1,1)$        & 0   &-1/2 &-3/4&1/2&3/2&0 & 0 & 0 & $(1,1)$ &-2 & 0\\
$\bar{S}_8$& $(1,1)$        & 0   &-1/2 &-3/4 &-1/2&-3/2& 0 & 0 & 0 & $(1,1)$ &-2 & 0\\
$\bar{S}_9$& $(1,1)$        & 0   &-1/2 & 5/4 &-1/2& 1/2& 0 & 0 & 0 & $(1,1)$ &-2 & 0\\
\hline
\end{tabular}
\vfill
\newpage

\no Table A.4: Hidden Sector Non-Abelian Fields
\begin{tabular}{|l||c|cccccccc|c|cc|}
\hline
  $F$      & $(SU(3)_C,$ & $Q_{Y}$ & $Q_{Z'}$
   & $Q_A$ & $Q_{1}$ & $Q_{2}$ & $Q_{3}$         & $Q_{4}$
           & $Q_{5}$ & $(SU(5)_{H},$   & $Q_{6}$ & $Q_{7}$  \\
           & $SU(2)_L)$ & & & & & & & & & $SU(3)_{H})$ & &  \\
\hline
      $F_1$& $(1,1)$        &-1/4 & 0   &-1/2 &-1/2 & 1/2& 0   & 0   & 0   & $(5,1)$ &-1 &-3\\
      $F_2$& $(1,1)$        & 1   & 0   & 0   & 0   & 1  & 0   & 0   & 1/2 & $(5,1)$ & 1 &-3\\
      $F_3$& $(1,1)$        & 1   & 0   & 0   &-1/2 &-1/2& 1/2 & 0   & 0   & $(5,1)$ & 1 &-3\\
      $F_4$& $(1,1)$        & 1   & 0   & 0   & 1/2 &-1/2& 0   &-1/2 & 0   & $(5,1)$ & 1 &-3\\
\hline
$\bar{F}^{'}_1$& $(1,1)$        & 1/4 & 0   & 1/2 & 1/2 &-1/2& 0   & 0   & 0   & $(\bar{5},1)$ & 1 & 3\\
$\bar{F}^{'}_2$&$(1,1)$     & 1   & 0   & 0   & 0   & 1  & 0   & 0   &-1/2 & $(\bar{5},1)$ &-1 & 3\\
$\bar{F}^{'}_3$&$(1,1)$     & 1   & 0   & 0   &-1/2 &-1/2&-1/2 & 0   & 0   & $(\bar{5},1)$ &-1 & 3\\
$\bar{F}^{'}_4$&$(1,1)$     & 1   & 0   & 0   & 1/2 &-1/2& 0   & 1/2 & 0   & $(\bar{5},1)$ &-1 & 3\\
\hline
      $K_1$& $(1,1)$        & 1   & 0   & 0   & 1/2 &-1/2& 0   & 1/2 & 0   & $(1,3)$ &-1 &-5\\
      $K_2$& $(1,1)$        & 1   & 0   & 0   &-1/2 &-1/2& 1/2 & 0   & 0   & $(1,3)$ &-1 &-5\\
      $K_3$& $(1,1)$        & 1   & 0   & 0   & 0   & 1  & 0   & 0   & 1/2 & $(1,3)$ &-1 &-5\\
      $K_4$& $(1,1)$        & 1/4 & 0   & 1/2 &-1/2 &-1/2& 0   & 0   & 0   & $(1,3)$ & 1 &-5\\
\hline
$\bar{K}^{'}_1$&$(1,1)$     & 1   & 0   & 0   & 1/2 &-1/2& 0   &-1/2 & 0   & $(1,\bar{3})$ & 1 & 5\\
$\bar{K}^{'}_2$&$(1,1)$     & 1   & 0   & 0   &-1/2 &-1/2&-1/2 & 0   & 0   & $(1,\bar{3})$ & 1 & 5\\
$\bar{K}^{'}_3$&$(1,1)$     & 1   & 0   & 0   & 0   & 1  & 0   & 0   &-1/2 & $(1,\bar{3})$ & 1 & 5\\
$\bar{K}_4$& $(1,1)$        &-1/4 & 0   &-1/2 & 1/2 & 1/2& 0   & 0   & 0   & $(1,\bar{3})$ & 1 & 5\\
\hline
\end{tabular}
\vfill
\end{flushleft}
\newpage

\section{Optical Unification Model Superpotential}

\def\ha{\half}
\begin{flushleft}
\no Table B.1 Hidden Sector and Singlet Terms to 6$^{\rm th}$ Order.\\
\no Coupling coefficients are given only for third order terms.
\beqn 
&&\mbox{3rd order:}\nolabel\\
&& \ha ( F_1 \bar{F}^{'}_1 \Phi_1 + 
\ha K_4 \bar{K}_4 \Phi_2 + 
S_1 \bar{S}_1 \Phi_3 +
S_2 \bar{S}_2 \Phi_3 +
S_3 \bar{S}_3 \Phi_3 +
S_4 \bar{S}_4 \Phi_3 +
S_5 \bar{S}_5 \Phi_3 +\nolabel\\&& 
S_6 \bar{S}_6 \Phi_3 +
S_7 \bar{S}_7 \Phi_1 +
S_8 \bar{S}_8 \Phi_1 +
S_9 \bar{S}_9 \Phi_1 ) +
S_1 S_2 \Phi_{12} +
S_3 S_4 \Phi_{12} + 
S_5 S_6 \Phi_{12} +\nolabel\\&&
S_7 \bar{S}_8 \Phi_{23} +
\bar{S}_1 \bar{S}_2 \bar{\Phi}_{12} +  
\bar{S}_3 \bar{S}_4 \bar{\Phi}_{12} +
\bar{S}_5 \bar{S}_6 \bar{\Phi}_{12} +
S_8 \bar{S}_7 \bar{\Phi}_{23} + 
\Phi_{12} \bar{\Phi}_{23} \bar{\Phi}_{31} +
\nolabel\\ 
&&\Phi_{23} \Phi_{31} \bar{\Phi}_{12}
\label{b13}\\
&&\nolabel\\  
&&\mbox{4th order:}\nolabel\\
&&F_1 \bar{F}^{'}_2 S_9 N_3 + 
F_3 \bar{F}^{'}_3 S_9 \bar{S}_7 + 
K_2 \bar{K}^{'}_2 S_9 \bar{S}_7  
\label{b14}\\
&&\nolabel\\  
&&\mbox{5th order:}\nolabel\\ 
&&
S_1 S_2 S_8 \bar{S}_7 \Phi_{31} + 
S_3 S_4 S_8 \bar{S}_7 \Phi_{31} + 
S_5 S_6 S_8 \bar{S}_7 \Phi_{31} +
S_7 \bar{S}_1 \bar{S}_2 \bar{S}_8 \bar{\Phi}_{31} + \nolabel\\
&&S_7 \bar{S}_3 \bar{S}_4 \bar{S}_8 \bar{\Phi}_{31} +
S_7 \bar{S}_5 \bar{S}_6 \bar{S}_8 \bar{\Phi}_{31}+
F_1 \bar{F}^{'}_3 S_9 \Phi_3 N_2 +
F_2 \bar{F}^{'}_2 S_9 \bar{S}_8 \Phi_2 +
\nolabel\\
&&K_3 \bar{K}^{'}_3 S_9 \bar{S}_8 \Phi_2  
\label{b15}\\
&&\nolabel\\
&&\mbox{6th order:} 
\nolabel\\&&
F_1 F_1 \bar{F}^{'}_1 \bar{F}^{'}_2 S_9 N_3 +
F_1 F_3 \bar{F}^{'}_1 \bar{F}^{'}_3 S_9 \bar{S}_7 +
F_1 \bar{F}^{'}_1 K_2 \bar{K}^{'}_2 S_9 \bar{S}_7 +
F_1 \bar{F}^{'}_2 K_4 \bar{K}_4 S_9 N_3 +
\nolabel\\&&
F_1 \bar{F}^{'}_2 S_9 N_3\sum_{i=1}^{9} S_i \bar{S}_i  +
F_1 \bar{F}^{'}_2 S_9 N_3 \Phi_{31} \bar{\Phi}_{31} +
F_1 \bar{F}^{'}_2 S_9 N_3 \Phi_2 \Phi_2 +
F_3 \bar{F}^{'}_3 S_9 \bar{S}_7 K_4 \bar{K}_4 + 
\nolabel\\&&
F_3 \bar{F}^{'}_3 S_9 \bar{S}_7 \sum_{i=1}^{9} S_i \bar{S}_i +
F_3 \bar{F}^{'}_3 S_9 \bar{S}_7 \Phi_{12} \bar{\Phi}_{12} +
F_3 \bar{F}^{'}_3 S_9 \bar{S}_7 \Phi_3 \Phi_3 +
F_4 \bar{F}^{'}_4 S_9 \bar{S}_7 S_1 S_2 +
\nolabel\\&&
F_4 \bar{F}^{'}_4 S_9 \bar{S}_7 S_3 S_4 +
F_4 \bar{F}^{'}_4 S_9 \bar{S}_7 S_5 S_6 +
K_1 \bar{K}_1   S_9 \bar{S}_7 S_1 S_2 +
K_1 \bar{K}^{'}_{1} S_9 \bar{S}_7  S_3 S_4 +
\nolabel\\&&
K_1 \bar{K}_1 S_9 \bar{S}_7 S_5 S_6 +
K_1 \bar{K}_4 S_4 S_9 \Phi_2 N_1 +
K_2 \bar{K}^{'}_{2} S_9 \bar{S}_7 K_4 \bar{K}_4 +
K_2 \bar{K}^{'}_{2} S_9 \bar{S}_7 \sum_{i=1}^{9} S_i \bar{S}_i +
\nolabel\\&&
K_2 \bar{K}^{'}_{2} S_9 \bar{S}_7 \Phi_{12} \bar{\Phi}_{12} +
K_2 \bar{K}^{'}_{2} S_9 \bar{S}_7 \Phi_3 \Phi_3 +
K_2 \bar{K}_4 S_1 S_9 \Phi_1 N_2
\label{b16}
\eeqn
\end{flushleft}
\vfill
\newpage

\section{Optical Unification Model Mass Matrices}

\no Table C.1a Possible Exotic Triplet $D\bD$ Mass Matrix to 6$^{\rm th}$ Order
\vskip 0.5truecm
\beqn
M_{D_i,\bar{D}_{j}}=
{\left(
\begin{array}{cccc}
\half \Phi_{3} + m^{'}_{11} & S_{9} \bar{S}_{8}\Phi_{2} &- & - \\
S_{8} \bar{S}_{9} + m^{'}_{21} & \half \Phi_{3} + m^{'}_{22} & - & \bar{S}_{9} + m^{'}_{24}\\
-    & -    &  K_{4} \bar{K}_{4} \sum_{i=1}^{6} S_i \bar{S}_i & - \\
S_{8} + m^{'}_{41} & - &-     & \half\Phi_{2} + m^{'}_{44}\\
\end{array} \right )}
\label{mijd1}
\eeqn
where,
\begin{flushleft}
\beqn 
m^{'}_{11} &=&
  F_{1} \bar{F}^{'}_{2} S_{9} N_{3}+ 
  F_{3} \bar{F}^{'}_{3 } S_{9} \bS_{7}+
  K_{2} \bar{K}^{'}_{2} S_{9} \bS_{7}
\label{dd11}\\
\nolabel\\ 
m^{'}_{21} &=& (S_{8} \bar{S}_{7}+  \Phi_{23} \Phi_{1} )(F_{3} \bar{F}^{'}_{3} + K_{2} \bar{K}^{'}_{2}) +
 \bar{\Phi}_{31} \Phi_{2} F_{4} \bar{F}^{'}_{4} + \bar{\Phi}_{31} \Phi_{2} K_{1} \bar{K}^{'}_{1}  
\label{dd21}\\
\nolabel\\ 
m^{'}_{22} &=& S_9 F_{1}\bar{F}^{'}_{2} N_{3} +
              S_{9} \bar{S}_{7} (F_{3}\bar{F}^{'}_{3} + K_{2}\bar{K}^{'}_{2})  
\label{dd22}\\
\nolabel\\ 
m^{'}_{24} &=& \bar{S}_9 F_1 \bar{F}_1 + \bar{S}_9 K_4 \bar{K}_4 + 
               \bar{S}_9 \sum_{i=1}^{9} S_{i}\bar{S}_{i} +
F_{1}\bar{F}^{'}_2 N_3 \Phi_{1} + 
\nolabel\\&& 
\bar{S}_{7} \Phi_{1}  (F_{3}\bar{F}^{'}_{3} + K_{2}\bar{K}^{'}_{2}) + 
\bar{S}_{7} \Phi_{12} (F_{4}\bar{F}^{'}_{4} + K_{2}\bar{K}^{'}_{2}) +
\nolabel\\&& 
\bar{S}_{8} \Phi_{23} (F_{3}\bar{F}^{'}_{3} + K_{2}\bar{K}^{'}_{2})  +
\bar{S}_{8} \Phi_{31} (F_{4}\bar{F}^{'}_{4} + K_{1}\bar{K}^{'}_{1})  +
\nolabel\\&& 
S_{1} N_{2} K_{2} \bar{K}_4 + S_{4} N_{1} K_{1} \bar{K}_4 
\label{dd24}\\
&&\nolabel\\ 
m^{'}_{41}&=& S_{8} F_{1}\bar{F}_{1} +
              S_{8} K_{4}\bar{K}_{4} + 
              S_{8} \sum_{i=7}^{9} S_{i}\bar{S}_{i}+
\nolabel\\&& 
S_{9} \Phi_{23} (F_{3}\bar{F}^{'}_{3} + K_{2}\bar{K}^{'}_{2} ) +
S_{9} \bar{\Phi}_{31} (F_{4}\bar{F}^{'}_{4} + K_{1}\bar{K}^{'}_{1}) 
\label{dd41}\\
&&\nolabel\\ 
m^{'}_{44}&=& 
S_9 N_{3} F_{1}\bar{F}^{'}_{2} +
S_{9}\bar{S}_{7} (F_{3}\bar{F}^{'}_{3} + K_{2}\bar{K}^{'}_{2})  
\label{dd44}
\eeqn
\end{flushleft}
\vfill
\newpage

\begin{flushleft}
{\no Table C.1b Possible Seventh Order $D \bD$ Mass Terms}    
\beqn
 D_{1}  \bD_{1}:&&   
       F_{1}   F_{1}  \bF_{1}  \bF_{1}   \P_{3}
 +     F_{1}  \bF_{1}   K_{4} \bK_{4}  \P_{1} 
 +     F_{1}  \bF_{1}   K_{4} \bK_{4}  \P_{2}
 +     F_{1}  \bF_{1}   K_{4} \bK_{4}  \P_{3}
 +     \nolabel\\&&
       F_{1}  \bF_{1}   \ss{1}{6}     \P_{1}
 +     F_{1}  \bF_{1}   \ss{7}{9}     \P_{3}
 +     F_{1}  \bF_{3}  S_{9}  \P_{3}  N_{2}
 +     F_{1}  \bF_{4}  S_{9}  \P_{3}  N_{1} 
 +     \nolabel\\&&
       F_{2} \bF_{2}  S_{9} \bS_{8}  \P_{2}
 +     K_{3} \bK_{3}  S_{9} \bS_{8}  \P_{2}
 +     K_{4}  K_{4} \bK_{4} \bK_{4}  \P_{3}
 +     K_{4} \bK_{4}  \ss{1}{6}     \P_{2}
 +     \nolabel\\&&
       K_{4} \bK_{4}  \ss{7}{9}     \P_{1} 
 +     K_{4} \bK_{4}  \ss{7}{9}     \P_{2}
 +     K_{4} \bK_{4}  \ss{7}{9}     \P_{3}
 +     \nolabel\\&&
       K_{4} \bK_{4}  S_{7} \bS_{8} \P_{23}
 +     K_{4} \bK_{4}  S_{8} \bS_{7} \bP_{23}
 +     S_{1}   S_{2}  S_{8} \bS_{7} \P_{31}
 +     \ss{1}{6}     \ss{7}{9}     \P_{1} 
 +     \nolabel\\&&
       \ss{1}{6}     S_{7} \bS_{8}  \P_{23}
 +     \ss{1}{6}     S_{8} \bS_{7} \bP_{23}
 +     S_{3}  S_{4}  S_{8} \bS_{7}  \P_{31}
 +     S_{5}  S_{6}  S_{8} \bS_{7}  \P_{31}
 +     \nolabel\\&&
       \ss{7}{9}     \ss{7}{9}     \P_{3}
 +      S_{7} \bS_{8} \bsso{1}{3} \bP_{31}
\label{x11}\\
 D_{1}  \bD_{2}:&&  
      F_{1}  \bF_{1}   S_{9} \bS_{8}  \P_{2}
 +    K_{4} \bK_{4}  S_{9} \bS_{8}  \P_{2}
 +    K_{4} \bK_{4}  S_{9} \bS_{7} \bP_{23}
 +    K_{4} \bK_{4}  S_{9} \bS_{8}  \P_{1} 
 +    \nolabel\\&&
      S_{9} \bS_{8}  \ss{1}{9}    \P_{2}  
 +    S_{9} \bS_{8}  \P_{31} \bP_{31}  \P_{2}
 +    S_{9} \bS_{8}  \P_{2}  \P_{2}  \P_{2}
\label{x12}\\
 D_{1}  \bD_{4}:&&  
    F_{1}  \bF_{1}   K_{4} \bK_{4} \bS_{8}
 +  F_{1}  \bF_{1}   \ss{1}{6}    \bS_{8}
 +  F_{1}  \bF_{1}  \bS_{8}  \P_{2}  \P_{3}
 +  F_{1}  \bF_{3}  S_{9} \bS_{8}  N_{2}
 +  \nolabel\\&&
    F_{2} \bF_{2}  S_{9} \bS_{8} \bS_{8}
 +  K_{3} \bK_{3}  S_{9} \bS_{8} \bS_{8}
 +  K_{4} \bK_{4}  \ss{1}{9}   \bS_{8}
 +  K_{4} \bK_{4} \bS_{7} \bP_{23} \P_{3}
 +  \nolabel\\&&
    K_{4} \bK_{4} \bS_{8}  \P_{1}   \P_{3}
 +  \ss{1}{6}   \ss{7}{9} \bS_{8}
 +  \ss{1}{6}    \bS_{7} \bP_{23} \P_{2}
 +  \ss{1}{6}    \bS_{8}  \P_{1}   \P_{2}
 +  \nolabel\\&&
    \sso{1}{3}   \bS_{7} \P_{31}  \P_{2}
 +  \ss{7}{9}    \bS_{8}  \P_{2}  \P_{3}
\label{x14}\\
 D_{2} \bD_{1}:&&
       F_{1}  \bF_{1}   F_{2} \bF_{2}  \P_{1} 
 +     F_{1}  \bF_{1}   K_{3} \bK_{3}  \P_{1} 
 +     F_{1}  \bF_{1}   S_{8} \bS_{9}  \P_{2}
 +     F_{1}  \bF_{3}  S_{7}  \P_{23} N_{2}
 +     \nolabel\\&&
       F_{1}  \bF_{3}  S_{8}  \P_{1}   N_{2}
 +     F_{1}  \bF_{3}  S_{8}  \P_{2}  N_{2}
 +     F_{1}  \bF_{3}  S_{8}  \P_{3}  N_{2}
 +     F_{1}  \bF_{4}  S_{8}  \P_{2}  N_{1} 
 +     \nolabel\\&&
       F_{2} \bF_{2}  K_{4} \bK_{4}  \P_{2}
 +     F_{2} \bF_{2}  S_{7} \bS_{8} \P_{23}
 +     F_{2} \bF_{2}  S_{8} \bS_{7} \bP_{23}
 +     F_{2} \bF_{2}  \ss{7}{9}     \P_{1} 
 +     \nolabel\\&&
       K_{3}  K_{4} \bK_{3} \bK_{4}  \P_{2}
 +     K_{3} \bK_{3}  \ss{7}{9}     \P_{1} 
 +     K_{3} \bK_{3}  S_{7} \bS_{8} \P_{23}
 +     K_{3} \bK_{3}  S_{8} \bS_{7} \bP_{23}
 +     \nolabel\\&&
       K_{3} \bK_{4}  S_{5}  S_{8}  N_{3}
 +     K_{4} \bK_{4}  S_{7} \bS_{9} \P_{23}
 +     K_{4} \bK_{4}  S_{8} \bS_{9}  \P_{2}
 +     S_{8} \bS_{9}  \ss{1}{9}      \P_{2}
 +     \nolabel\\&&
       S_{8} \bS_{9} \P_{31} \bP_{31}  \P_{2}
 +     S_{8} \bS_{9}  \P_{2}  \P_{2}  \P_{2}
\label{x21}\\
    D_{2} \bD_{2}:&&
      F_{1}   F_{1}  \bF_{1}  \bF_{1}   \P_{3}
 +    F_{1}  \bF_{1}   K_{4} \bK_{4}  \P_{1} 
 +    F_{1}  \bF_{1}   K_{4} \bK_{4}  \P_{2}
 +    F_{1}  \bF_{1}   K_{4} \bK_{4}  \P_{3}
 +    \nolabel\\&&
      F_{1}  \bF_{1}  \ss{1}{6}      \P_{1} 
 +    F_{1}  \bF_{1}  \ss{7}{9}      \P_{3}
 +    F_{1}  \bF_{3}  S_{9}  \P_{3}  N_{2}
 +    F_{1}  \bF_{4}  S_{9}  \P_{3}  N_{1} 
 +    \nolabel\\&&
      F_{2} \bF_{2}  S_{9} \bS_{8}  \P_{2}
 +    K_{3} \bK_{3}  S_{9} \bS_{8}  \P_{2}
 +    K_{4}  K_{4} \bK_{4} \bK_{4}  \P_{3}
 +    K_{4} \bK_{4} \ss{1}{9}      \P_{2}
 +    \nolabel\\&&
      K_{4} \bK_{4} \ss{7}{9}      \P_{1} 
 +    K_{4} \bK_{4} \ss{7}{9}      \P_{3}
 +    K_{4} \bK_{4}  S_{7} \bS_{8} \P_{23}
 +    K_{4} \bK_{4}  S_{8} \bS_{7} \bP_{23}
 +    \nolabel\\&&
      S_{1}   S_{2}  S_{8} \bS_{7} \P_{31}
 +    \ss{1}{6}     \ss{7}{9}     \P_{1} 
 +    \ss{1}{6}     S_{7} \bS_{8} \P_{23}
 +    \ss{1}{6}     S_{8} \bS_{7} \bP_{23}
 +    \nolabel\\&&
      S_{3}  S_{4}  S_{8} \bS_{7} \P_{31}
 +    S_{5}  S_{6}  S_{8} \bS_{7} \P_{31}
 +    \ss{7}{9}     \ss{7}{9}     \P_{3}
 +    \nolabel\\&&
      S_{7} \bS_{8} \bsso{1}{3}  \bP_{31}
\label{x22}\\
 D_{2} \bD_{4}:&&
      F_{1}   F_{1}  \bF_{1}  \bF_{3}  N_{2}
 +    F_{1}   F_{1}  \bF_{1}  \bF_{4}  N_{1} 
 +    F_{1}   F_{2} \bF_{1}  \bF_{2} \bS_{8}
 +    F_{1}  \bF_{1}   K_{3} \bK_{3} \bS_{8}
 +    \nolabel\\&&
      F_{1}  \bF_{1}  \bS_{9}  F_{1}  \bF_{1}   
 +    F_{1}  \bF_{1}  \bS_{9}  K_{4} \bK_{4}  
 +    F_{1}  \bF_{1}  \bS_{9}  \ss{1}{9}    
 +    F_{1}  \bF_{1}  \bS_{9} \P_{12}\bP_{12}
 +    \nolabel\\&&
      F_{1}  \bF_{1}  \bS_{9} \P_{31} \bP_{31}
 +    F_{1}  \bF_{1}  \bS_{9}  \P_{2}  \P_{2}
 +    F_{1}  \bF_{1}  \bS_{9}  \P_{3}  \P_{3}
 +    F_{1}  \bF_{3} \ss{1}{9}     N_{2}
 +    \nolabel\\&&
      F_{1}  \bF_{3}  \P_{1}   \P_{3}  N_{2}
 +    F_{1}  \bF_{4}  K_{4} \bK_{4}  N_{1} 
 +    F_{1}  \bF_{4} \ss{1}{9}     N_{1} 
 +    F_{1}  \bF_{4}  \P_{12}\bP_{12} N_{1} 
 +    \nolabel\\&&
      F_{1}  \bF_{4} \P_{31} \bP_{31}  N_{1} 
 +    F_{1}  \bF_{4}  \P_{2}  \P_{2}  N_{1} 
 +    F_{1}  \bF_{4}  \P_{3}  \P_{3}  N_{1} 
 +    F_{2} \bF_{2}  K_{4} \bK_{4} \bS_{8}
 +    \nolabel\\&&
      F_{2} \bF_{2}  \ss{7}{9}    \bS_{8}
 +    F_{2} \bF_{2} \bS_{7} \bP_{23} \P_{2}
 +    F_{2} \bF_{2} \bS_{8}  \P_{1}   \P_{2}
 +    K_{3}  K_{4} \bK_{3} \bK_{4} \bS_{8}
 +    \nolabel\\&&
      K_{3} \bK_{3} \ss{7}{9}     \bS_{8}
 +    K_{3} \bK_{3} \bS_{7} \bP_{23} \P_{2}
 +    K_{3} \bK_{3} \bS_{8}  \P_{1}   \P_{2}
 +    K_{3} \bK_{4}  S_{5}  \P_{2}  N_{3}
 +    \nolabel\\&&
      K_{4}  K_{4} \bK_{4} \bK_{4} \bS_{9}
 +    K_{4} \bK_{4} \ss{1}{9}    \bS_{9}
 +    K_{4} \bK_{4} \bS_{9} \P_{12}\bP_{12}
 +    K_{4} \bK_{4} \bS_{9} \P_{23}\bP_{23}
 +    \nolabel\\&&
      K_{4} \bK_{4} \bS_{9}  \P_{1}   \P_{1} 
 +    K_{4} \bK_{4} \bS_{9}  \P_{3}  \P_{3}
 +    \ss{1}{9}\ss{1}{9}       \bS_{9}
 +    \ss{1}{6}    \bS_{9} \P_{23}\bP_{23}
 +    \nolabel\\&&
      \ss{1}{6}    \bS_{9} \P_{31} \bP_{31}
 +    \ss{1}{6}    \bS_{9}  \P_{1}   \P_{1} 
 +    \ss{1}{9}   \bS_{9}  \P_{2}  \P_{2}
 +    \ss{7}{9}    \bS_{9}  \P_{3}  \P_{3}
 +    \nolabel\\&&
      \ss{7}{9}    \bS_{9} \P_{12}\bP_{12}
 +    S_{1}   S_{2} \bS_{3} \bS_{4} \bS_{9}
 +    S_{5}  S_{6} \bS_{3} \bS_{4} \bS_{9}
 +    S_{3}  S_{4} \bS_{1}  \bS_{2} \bS_{9}
 +    \nolabel\\&&
      S_{5}  S_{6} \bS_{1}  \bS_{2} \bS_{9}
 +    S_{3}  S_{4} \bS_{5} \bS_{6} \bS_{9}
 +    \sso{1}{3}    \bS_{9} \P_{23} P_{31}
 +    S_{7} \bS_{8} \bS_{9} \P_{12}\bP_{31}
 +    \nolabel\\&&
      S_{8} \bS_{7} \bS_{9} \P_{31} \bP_{12}
 +    \ss{7}{9}    \bS_{9}  \P_{31} \bP_{31}
 +    \bsso{1}{3}    \bS_{9} \bP_{23}\bP_{31}
\label{x24}\\
 D_{4} \bD_{1}:&&   
       F_{1}   F_{1}  \bF_{1}  \bF_{1}   S_{8}
 +     F_{1}   F_{2} \bF_{1}  \bF_{2}  S_{9}
 +     F_{1}  \bF_{1}   K_{3} \bK_{3}  S_{9}
 +     F_{1}  \bF_{1}   K_{4} \bK_{4}  S_{8}
 +     \nolabel\\&&
       F_{1}  \bF_{1}   \ss{1}{9}    S_{8}  
 +     F_{1}  \bF_{1}   S_{7}  \P_{12}\bP_{31}
 +     F_{1}  \bF_{1}   S_{8} \P_{12}\bP_{12}
 +     F_{1}  \bF_{1}   S_{8} \P_{31} \bP_{31}
 +     \nolabel\\&&
       F_{1}  \bF_{1}   S_{8}  \P_{2}  \P_{2}
 +     F_{1}  \bF_{1}   S_{8}  \P_{3}  \P_{3}
 +     F_{1}  \bF_{3}  S_{8}  S_{9}  N_{2}
 +     F_{1}  \bF_{4}  S_{8}  S_{9}  N_{1} 
 +     \nolabel\\&&
       F_{2} \bF_{2}  K_{4} \bK_{4}  S_{9}
 +     F_{2} \bF_{2}  S_{7}  S_{9} \bS_{7}
 +     F_{2} \bF_{2}  S_{8}  S_{9} \bS_{8}
 +     F_{2} \bF_{2}  S_{9}  S_{9} \bS_{9}
 +     \nolabel\\&&
       F_{2} \bF_{2}  S_{9}  \P_{1}   \P_{2}
 +     K_{3} \bK_{3}  S_{9}  K_{4} \bK_{4}   
 +     K_{3} \bK_{3}  S_{7}  S_{9} \bS_{7}
 +     K_{3} \bK_{3}  S_{8}  S_{9} \bS_{8}
 +     \nolabel\\&&
       K_{3} \bK_{3}  S_{9}  S_{9} \bS_{9}
 +     K_{3} \bK_{3}  S_{9}  \P_{1}   \P_{2}
 +     K_{4} \bK_{4}  S_{8}  K_{4} \bK_{4}   
 +     K_{4} \bK_{4}  S_{8}  \ss{1}{9}       
 +     \nolabel\\&&
       K_{4} \bK_{4}  S_{8}  \P_{12}\bP_{12}
 +     K_{4} \bK_{4}  S_{8}  \P_{23}\bP_{23}
 +     K_{4} \bK_{4}  S_{8}  \P_{1}   \P_{1} 
 +     K_{4} \bK_{4}  S_{8}  \P_{3}  \P_{3}
 +     \nolabel\\&&
       \ss{1}{9}    S_{8} \ss{1}{9}       
 +     S_{1}   S_{2}  S_{8} \bS_{3} \bS_{4}
 +     S_{1}   S_{2}  S_{8} \bS_{5} \bS_{6}
 +     S_{3}  S_{4}  S_{8} \bS_{1}  \bS_{2}
 +     \nolabel\\&&
       S_{3}  S_{4}  S_{8} \bS_{5} \bS_{6}
 +     S_{5}  S_{6}  S_{8} \bS_{1}  \bS_{2}
 +     S_{5}  S_{6}  S_{8} \bS_{3} \bS_{4}
 +     \sso{1}{3}     S_{8} \P_{23} P_{31}
 +     \nolabel\\&&
       \ss{1}{6}     S_{8} \P_{23}\bP_{23}
 +     \ss{1}{9}    S_{8}  \P_{31} \bP_{31}
 +     K_{4} \bK_{4}  S_{7}  \P_{23} \P_{1} 
 +     \ss{1}{6}     S_{7}  \P_{23} \P_{1} 
 +     \nolabel\\&&
       \ss{1}{6}     S_{8}  \P_{1}   \P_{1} 
 +     \ss{1}{6}     S_{8}  \P_{2}  \P_{2}
 +     \ss{7}{9}     S_{7}  \P_{12}\bP_{31}     
 +     \ss{7}{9}     S_{8}  \P_{12}\bP_{12}
 +     \nolabel\\&&
       \bsso{1}{3}    S_{7}  \P_{31}  \P_{1} 
 +     S_{8} \bS_{7}  S_{8}  \P_{31} \bP_{12}
 +     \ss{7}{9}     S_{8}  \P_{3}  \P_{3}
 +     \nolabel\\&&
       \bsso{1}{3}    S_{8} \bP_{23}\bP_{31}
\label{x41}\\
    D_{4} \bD_{2}:&&
      F_{1}  \bF_{1}   K_{4} \bK_{4}  S_{9}
 +    F_{1}  \bF_{1}   \ss{1}{6}     S_{9} 
 +    F_{1}  \bF_{1}   S_{9}  \P_{2}  \P_{3}
 +    F_{1}  \bF_{3}  S_{9}  S_{9}  N_{2}
 +    \nolabel\\&&
      F_{2} \bF_{2}  S_{9}  S_{9} \bS_{8}
 +    K_{3} \bK_{3}  S_{9}  S_{9} \bS_{8}
 +    K_{4} \bK_{4}  \ss{1}{9}    S_{9} 
 +    K_{4} \bK_{4}  S_{9}  \P_{1}   \P_{3}
 +    \nolabel\\&&
      \ss{1}{6}     \ss{7}{9}     S_{9} 
 +    \ss{1}{9}    S_{9}  \P_{1}   \P_{2}
\label{x42}\\
    D_{4} \bD_{4}:&&
      F_{1}  \bF_{1}   F_{1}  \bF_{1}   \P_{2}
 +    F_{1}  \bF_{1}   K_{4} \bK_{4}  \P_{1} 
 +    F_{1}  \bF_{1}   \sso{1}{3}     \P_{12}
 +    F_{1}  \bF_{1}   \ss{1}{6}     \P_{1} 
 +    \nolabel\\&&
      F_{1}  \bF_{1}   \ss{1}{6}     \P_{2}
 +    F_{1}  \bF_{1}   \ss{1}{6}     \P_{3}
 +    F_{1}  \bF_{1}   \ss{7}{9}     \P_{2}
 +    F_{1}  \bF_{1}  \bsso{1}{3}    \bP_{12}
 +    \nolabel\\&&
      F_{1}  \bF_{3}  S_{9}  \P_{3}  N_{2}
 +    F_{1}  \bF_{4}  S_{9}  \P_{2}  N_{1} 
 +    F_{2} \bF_{2}  S_{9} \bS_{7} \bP_{23}
 +    K_{3} \bK_{3}  S_{9} \bS_{7} \bP_{23}
 +    \nolabel\\&&
      F_{2} \bF_{2}  S_{9} \bS_{8}  \P_{1} 
 +    K_{3} \bK_{3}  S_{9} \bS_{8}  \P_{1} 
 +    F_{2} \bF_{2}  S_{9} \bS_{8}  \P_{2}
 +    K_{3} \bK_{3}  S_{9} \bS_{8}  \P_{2}
 +    \nolabel\\&&
      K_{3} \bK_{4}  S_{5}  S_{9}  N_{3}
 +    K_{4} \bK_{4}  \sso{1}{3}     \P_{12}
 +    K_{4} \bK_{4}  \ss{1}{9}    \P_{3}
 +    K_{4} \bK_{4}  S_{7} \bS_{8}  \P_{23}
 +    \nolabel\\&&
      K_{4} \bK_{4}  S_{8} \bS_{7} \bP_{23}
 +    K_{4} \bK_{4} \bsso{1}{3}    \bP_{12}
 +    \sso{1}{3}     \ss{7}{9}     \P_{12}
 +    \nolabel\\&&
      \ss{1}{6}   \ss{7}{9}       \P_{1} 
 +    \ss{1}{6}   \ss{7}{9}       \P_{3}
 +    \ss{1}{9}   \ss{1}{9}     \P_{2}
 +    S_{1}   S_{2} \bS_{3} \bS_{4}  \P_{2}
 +    \nolabel\\&&
      S_{1}   S_{2} \bS_{5} \bS_{6}  \P_{2}
 +    S_{3}  S_{4} \bS_{1}  \bS_{2}  \P_{2}
 +    S_{3}  S_{4} \bS_{5} \bS_{6}  \P_{2}
 +    S_{5}  S_{6} \bS_{1}  \bS_{2}  \P_{2}
 +    \nolabel\\&&
      S_{5}  S_{6} \bS_{3} \bS_{4}  \P_{2}
 +    \sso{1}{3}     S_{8} \bS_{7}  \P_{31}
 +    \ss{1}{6}     S_{7} \bS_{8}  \P_{23}
 +    \ss{1}{6}     S_{8} \bS_{7} \bP_{23}
 +    \nolabel\\&&
      S_{7} \bS_{8}\bsso{1}{3}     \bP_{31}
 +    \ss{7}{9}  \bsso{1}{3}      \bP_{12}
\label{x44}
\eeqn
\end{flushleft}
\vfill
\newpage

\begin{flushleft}
\no Table C.2a Possible Exotic Doublet $X \bX$ Mass Matrix to 6$^{\rm th}$ Order
\beqn
M_{X_i,\bar{X}_{j}} = {\left(
\begin{array}{ccc}
(S_{1} \bar{S}_{6} + S_{5} \bar{S}_{2}) \Phi_{2}&-&-\\
&&\\
-&(S_{4} \bar{S}_{6} + S_{5} \bar{S}_{3}) \Phi_{1}&-\\
&&\\
-&-&-\\
\end{array} \right )}
\label{mijx}
\eeqn
\end{flushleft}

\begin{flushleft}
{\no Table C.2b Possible Seventh Order $\bX\bX$ and $X\bX$ Mass Terms}    
\beqn
\bX_{1} \bX_{3}:&&    \bK_{1}  \bK_{3} \bK_{4}  S_{1}  \bS_{8}
\label{x713m}\\
X_{1} \bX_{1}:&&    K_{4} \bK_{4}  S_{1}   S_{5}  \P_{12}
                    + K_{4} \bK_{4} \bS_{2} \bS_{6} \bP_{12}
                    + S_{1}   S_{5} \bS_{3} \bS_{4}  \P_{2}
                    + S_{3}  S_{4} \bS_{2} \bS_{6}  \P_{2}
\label{x711}\\
 X_{2} \bX_{2}:&&   F_{1}  \bF_{1}   S_{4}  S_{5}  \P_{12}
  +  F_{1}  \bF_{1}  \bS_{3} \bS_{6} \bP_{12}
  +  S_{1}   S_{2} \bS_{3} \bS_{6}  \P_{1} 
  +  S_{4}  S_{5} \bS_{1}  \bS_{2}  \P_{1}+ 
\nolabel\\&&
     S_{4}  S_{5} \ss{7}{9}      \P_{12}
  +  S_{4}  S_{5}  S_{8} \bS_{7}  \P_{31}
  +  S_{7} \bS_{3} \bS_{6} \bS_{8} \bP_{31}
  +  \nolabel\\&&
     \bS_{3}\bS_{6} \ss{7}{9}   \bP_{12}
\label{x722}\\ 
  X_{3} \bX_{3}:&&  F_{1}  \bF_{1}   S_{4} \bS_{2}  \P_{1} 
  +  F_{1}  \bF_{1}   S_{1}  \bS_{3}  \P_{1} 
  +  \ss{7}{9}    S_{4} \bS_{2}  \P_{1} 
  +  \ss{7}{9}    S_{1}  \bS_{3}  \P_{1} + 
\nolabel\\&&
    K_{4} \bK_{4}  S_{1}  \bS_{3}  \P_{2}
  +  K_{4} \bK_{4}  S_{4} \bS_{2}  \P_{2}
  +  S_{1}   S_{4}  S_{8} \bS_{7}  \P_{31}
  +  S_{1}   S_{7} \bS_{3} \bS_{8}  \P_{23} +
\nolabel\\&&
     S_{1}  S_{8} \bS_{3} \bS_{7} \bP_{23}
  +  S_{4}  S_{7} \bS_{2} \bS_{8}  \P_{23}
  +  S_{4}  S_{7} \bS_{2} \bS_{7}  \P_{1} 
  +  S_{4}  S_{8} \bS_{2} \bS_{7} \bP_{23} +
\nolabel\\&&
    S_{7} \bS_{2} \bS_{3} \bS_{8} \bP_{31}
\label{x733}
\eeqn
\end{flushleft}
\vfill
\newpage

\no Table C.3a Possible Exotic Hypercharged Singlet $A\bA$ Mass Matrix to 6$^{\rm th}$ Order
\begin{flushleft}
\beqn
M_{A_i,\bar{A}_{j}} &=&
{\left(
\begin{array}{ccccccc}
M_{11}&-&-&-&-&-&-\\
-& M_{22}&-&-&-&-&-\\
-&-& -&-&-&-&-\\
-&-&-& M_{44}&-&-&-\\
N_2 &-&-&-& M_{55}&-&-\\
-&-&-&-&-&-&-\\
M_{71}&-&-&-&-&-& M_{77}
\end{array} \right )}
\label{mija}
\eeqn
\no where,
\beqn
M_{11} &=& F_1 \bar{F}^{'}_{2} S_{9} N_{3}  
\label{a11}\\
\nolabel\\
M_{22} &=& M_{55} = (S_{3} \bar{S}_{5} + S_{6} \bar{S}_{4}) \Phi_{1} 
\label{a22}\\
\nolabel\\
M_{44} &=& M_{77} = (S_{2} \bar{S}_{5} + S_{6} \bar{S}_{1}) \Phi_{2} 
\label{a44}\\
\nolabel\\
M_{51} &=& N_2 
\label{a51}\\
\nolabel\\
M_{71} &=& (S_{2} \bar{S}_{3} + S_{4} \bar{S}_{1}) N_{1} \Phi_{2}
\label{a71}\\
\nolabel
\eeqn
\end{flushleft}
\vfill  
\newpage

\begin{flushleft}
{\no Table C.3b Possible Seventh Order $A A$ and $A\bA$ Mass Terms}    
\beqn
A_{3} A_{5}:&&   K_{2} K_{3} K_{4} S_{9} \bar{S}_{4} 
\label{b35}\\ 
A_{3} A_{6}:&&   K_{1} K_{1} K_{4} S_{2} S_{9} 
               + K_{2} K_{2} K_{4} S_{9} \bar{S}_{4} 
\label{b36}\\ 
A_{4} A_{6}:&&   K_{1} K_{3} K_{4} S_{2} S_{9}       
\label{b46}\\
A_{1} \bar{A}_{1}:&&
(F_{2} \bar{F}^{'}_{2} + K_{3} \bar{K}^{'}_{3}) S_{9}\bS_{8} \Phi_{2}+ 
( S_{1} S_{2} + S_{3} S_{4} + S_{5} S_{6})S_{8} \bar{S}_{7} \Phi_{31} +
\nolabel\\&&
(\bar{S}_{1} \bar{S}_{2} + \bar{S}_{3} \bar{S}_{4} + 
 \bar{S}_{5} \bar{S}_{6}) S_{7} \bar{S}_{8} \bar{\Phi}_{31} +
\label{b11}\\
A_{1} \bar{A}_{5}:&&   
F_{1} \bar{F}^{'}_{3} (S_{3} \bS_{5} + S_{6} \bS_{4}) S_{9}  
\label{c15}\\
A_{1} \bar{A}_{6}:&&   
F_{1} \bar{F}^{'}_{2} (S_{3} \bar{S}_{5}+ S_{6} \bar{S}_{4}) S_{9} 
\label{c16}\\
A_{1} \bar{A}_{7}:&&   
K_{1} \bar{K}_{4} S_{6} S_{9} \Phi_{2} 
\label{b17}\\
A_{2} \bar{A}_{2},\ A_{5} \bar{A}_{5}:&& 
F_{1} \bar{F}_{1} (S_{3} S_{6}\Phi_{12} + \bar{S}_{4} \bar{S}_{5} \bar{\Phi}_{12}) +
K_{3} \bar{K}_{4} N_{2} S_{3} S_{9} +
\nolabel\\&&
S_1 S_2 \bS_4 \bS_5 \Phi_{1} + S_3 S_6 \bS_1 \bS_2 \Phi_{1}+  
S_3 S_6 \sum_{i=7}^{9}S_{i}\bS_{i}\Phi_{12} + 
\nolabel\\&&
S_3 S_6 S_8 \bar{S}_7 \Phi_{31}+
\bar{S}_4 \bar{S}_5 S_7 \bar{S}_8 \bar{\Phi}_{31}+
\bar{S}_4 \bar{S}_5 \sum_{i=7}^{9} S_i \bar{S}_i \bar{\Phi}_{12} 
\label{b22}\\
A_{2} \bar{A}_{3}:&& K_{3} \bar{K}_{4} N_{2} S_{3} S_{9}  
\label{b23}\\
A_{2} \bar{A}_{4},\, A_{5} \bar{A}_{7}:&&
K_{1} \bar{K}_{4} N_{2} S_{6} S_{9}+ K_{2} \bar{K}_{4} N_{1} S_{9} \bar{S}_{5}  
\label{b24}\\
A_{3} \bar{A}_{3},\, A_{6} \bar{A}_{6}:&& 
F_{1} \bF_{1} (S_{2} \bS_{4} + S_{3} \bS_{1} )\Phi_{1}+
K_{4} \bar{K}_{4} (S_{2} \bS_{4} + S_{3} \bS_{1} )\Phi_{2}+
\nolabel\\&&
S_2 S_7 \bS_4 S_8 \Phi_{23} + 
S_3 S_7 \bS_1 S_8 \Phi_{23} + 
S_2 S_8 \bS_4 S_7 \bPhi_{23} + 
\nolabel\\&&
S_3 S_8 \bS_1 S_7 \bPhi_{23} + 
S_3 \bS_{1} \sum_{i=7}^{9} S_i \bS_i \Phi_{1}
\label{b33}\\
A_{3} \bar{A}_{4}:&& K_{1} \bar{K}_{4} N_{3} S_{2} S_{9}
\label{b34}\\
A_{4} \bar{A}_{2},\, A_{7} \bar{A}_{5}:&& K_{2} \bar{K}_{4} N_{1} S_{6} S_{9}
\label{b42}\\
A_{4} \bar{A}_{4},\, A_{7} \bar{A}_{7}:&& 
K_{3} \bar{K}_{4}  N_3 S_{2} S_{9} +
K_{4} \bar{K}_{4}  S_{2} S_{6} \Phi_{12} +
K_{4} \bar{K}_{4}  \bar{S}_{1} \bar{S}_{5}\bar{\Phi}_{12}+
\nolabel\\&&
S_{2} S_{6} \bS_{3} \bS_{4} \Phi_{2}+
S_{3} S_{4} \bS_{1} \bS_{5} \Phi_{2}
\label{b44}\\
A_{5} \bar{A}_{1}:&&  (F_{3} \bar{F}^{'}_{2}) N_3 S_9 \bar{S}_7 
\label{b51}\\
A_{6} \bar{A}_{1}:&& K_{2} \bar{K}^{'}_{3} N_{2} S_{9} \bar{S}_{8}
\label{b61}\\
A_{6} \bar{A}_{5}:&&  K_2 \bar{K}_4 N_3 S_3 S_9
\label{b65}\\
A_{7} \bar{A}_{6}:&&  K_3 \bar{K}_4 N_1 S_2 S_9
\label{b76}
\eeqn
\end{flushleft}
\vfill

\newpage

\no Table C.4a Possible Higgs Mass Matrix to 6$^{\rm th}$ Order
\vskip 0.3truecm
\beqn
M_{h_i,\bh{j}}= 
{\left(
\begin{array}{cccc}
-
&\Phi_{12} 
&\Phi_{31} 
& S_{9} 
\\
\\
\bar{\Phi}_{12} 
&m_{22}
&\bar{\Phi}_{23} 
&m_{24}
\\
\\
\bar{\Phi}_{31} 
&\Phi_{23}
&m_{33}
& -  
\\
\\
\bar{S}_9 + m_{41}  
&m_{42}&m_{43}
&\Phi_1 + m_{44}
\end{array} \right )}\label{mijh1}
\eeqn
where,
\beqn
m_{22} &=& F_{1}\bar{F}^{'}_{2}N_{3}S_{9} 
\label{h22}\\
m_{24} &=& (S_{1}S_{2}+S_{3}S_{4}+S_{5}S_{6}) S_{9}  
\label{h24}\\
m_{33} &=& (F_{3}\bar{F}^{'}_{3}+K_{2}\bar{K}^{'}_{2})S_{9}\bar{S}_{7}
\label{h33}\\
m_{41} &=& K_{2} \bar{K}_{4} N_{2} S_{1}
\label{h41}\\
m_{42} &=& ( \bar{S}_{1}\bar{S}_{2}+\bar{S}_{3}\bar{S}_{4}
           + \bar{S}_{5}\bar{S}_{6})\bar{S}_{9}
\label{h42}\\
m_{43} &=& 
(F_{3}\bar{F}^{'}_{3}+ K_{2} \bar{K}^{'}_{2})\bar{S}_{8}\Phi_{12} 
\label{h43}\\
m_{44} &=& F_{1}\bar{F}^{'}_{2} N_{3}S_{9} +
           (F_{3}\bar{F}^{'}_{3} + 
            K_{2}\bar{K}^{'}_{2})S_{9}\bar{S}_{7}
\label{h44}
\eeqn
\newpage

\def\phl{\phantom{i}}
\begin{flushleft}
\no Table C.4b Possible Seventh Order $h_{i}  \bar{h}_{j}$ Mass Terms

\beqn
h_2 \bar{h}_2:&&  
    F_2  \bF_2   S_9  \bS_8   \Phi_{2}   +
    K_3  \bK_3   S_9  \bS_8   \Phi_{2}   +
    S_1   S_2   S_8  \bS_7   \Phi_{31}  +
    S_3   S_4   S_8  \bS_7   \Phi_{31}  +
\nolabel\\&&
    S_5   S_6   S_8  \bS_7   \Phi_{31}  +
    S_7  \bS_1  \bS_2  \bS_8  \bPhi_{31}  +
    S_7  \bS_3  \bS_4  \bS_8  \bP_{31}  +
    S_7  \bS_5  \bS_6  \bS_8  \bP_{31}
\label{h2h2}\\ 
&&\nolabel\\
h_{3}  \bar{h}_{3}:&&  F_1  \bF_3   S_9   \Phi_{3}   N_2
\label{h3h3}\\ 
&&\nolabel\\
h_{3}  \bar{h}_{4}:&&
     (S_{1} S_{2} +  S_{3}   S_{4} + S_{5}   S_{6}) S_{8}   S_{9}  \bar{S}_{7} 
\label{h3h4}\\ &&\nolabel\\
h_{4}  \bar{h}_{1}:&&
      F_{1} \bar{F}^{'}_{3} N_{2}\sum_{i=1}^{6} S_{i} \bar{S}_{i}+
      F_{2}  \bar{F}^{'}_{2} K_{4}  \bar{K}_{4} \bS_8 +  
      K_{3} \bar{K}^{'}_{3}  K_{4}  \bar{K}_{4} \bS_8 +
\nolabel\\&&
      K_{3} \bar{K}_{4} N_3 S_5 \Phi_{2} 
\label{h4h1}\\
&&\nolabel\\
h_{4}  \bar{h}_{2}:&&
F_{1} \bF_{1} \bS_9 (\bS_1 \bS_2 + \bS_3 \bS_4 +\bS_5 \bS_6 )+ 
F_{1} \bF^{'}_{3} N_2 (\bS_1 \bS_2 + \bS_3 \bS_4 +\bS_5 \bS_6 )+ 
\nolabel\\&&
F_{1} \bF^{'}_{4} N_1 (\bS_1 \bS_2 + \bS_3 \bS_4 +\bS_5 \bS_6 )+ 
K_{4} \bK_{4} \bS_9 (\bS_1 \bS_2 + \bS_3 \bS_4 +\bS_5 \bS_6 )+ 
\nolabel\\&&
(F_2 \bF_2 + K_3 \bK_3)\bS_7 \Phi_{31} \Phi_{2} +
K_3 \bK_4 \bS_6 N_3 \Phi_{2} +
\sum_{i=1}^{9} (\bS_1 \bS_2 + \bS_3 \bS_4 +\bS_5 \bS_6 )+ 
\nolabel\\&&
\bS_9(\Phi_{31}\Phi_{31} + \Phi_{2}\Phi_{2})
(\bS_1 \bS_2 + \bS_3 \bS_4 +\bS_5 \bS_6 ) 
\label{h4h2}\\&&\nolabel\\
h_{4}  \bar{h}_{3}:&&
(\bar{S}_{1} \bar{S}_{2} + \bar{S}_{3}\bar{S}_{4} +\bar{S}_{5}\bar{S}_{6}) 
\bar{S}_{8}\bar{S}_{9}
\label{h4h3}\\&&\nolabel\\
h_{4}  \bar{h}_{4}:&&
 F_{1}  \bar{F}_{1} K_{4}  \bar{K}_{4} \Phi_{2} +
 F_{1}  \bar{F}^{'}_{3} N_2 S_9 \Phi_{3} +
 (F_{2}  \bar{F}^{'}_{2} + K_3 \bK^{'}_{3} )S_9 \bS_8 \Phi_{2} +
\nolabel\\&&
K_3 \bK_4 N_3 S_5 S_9 +
F_{1} \bar{F}_{1} 
(S_{1} S_{2}+ S_{3} S_{4}+ S_{5} S_{6}) \Phi_{12} +
F_{1} \bar{F}_{1} 
(\bar{S}_{1} \bar{S}_{2}+ 
 \bar{S}_{3} \bar{S}_{4}+
\nolabel\\&&
 \bar{S}_{5} \bar{S}_{6})\bar{\Phi}_{12}+
F_{1} \bar{F}_{1} 
(\bar{S}_{1} \bar{S}_{2}+ 
 \bar{S}_{3} \bar{S}_{4}+
 \bar{S}_{5} \bar{S}_{6})\bar{\Phi}_{12}+
\nolabel\\&&
F_{1} \bar{F}_{1} \sum_{i=1}^{6}S_i \bar{S}_{i}\Phi_{3}+ 
K_{4} \bar{K}_{4} \sum_{i=1}^{6}S_i \bar{S}_{i}(\Phi_{2}+\Phi_{3})+ 
\nolabel\\&&   
\sum_{i=7}^{9} S_i \bar{S}_{i} (S_{1} S_{2}+ S_{3} S_{4}+ S_{5} S_{6})+
\sum_{i=7}^{9} S_i \bar{S}_{i} (\bS_{1} \bS_{2}+ \bS_{3} \bS_{4}+ \bS_{5} \bS_{6})+
\nolabel\\&&
S_{8} \bar{S}_{7}\Phi_{31}+
(S_{1} S_{2}+ S_{3} S_{4} + S_{5} S_{6})+ 
S_{7} \bar{S}_{8}\Phi_{31}+
    (\bS_{1} \bS_{2}+ \bS_{3} \bS_{4} + \bS_{5} \bS_{6})+ 
\nolabel\\&&
\sum_{i=7}^{9} S_i \bar{S}_{i} \sum_{i=1}^{9} S_i \bar{S}_{i} 
\label{h4h4}
\eeqn
\end{flushleft}
\newpage

\no Table C.5 Possible Up-Quark Mass Matrix to 6$^{\rm th}$ Order
\vskip 0.3truecm
\beqn
M_{Q_i, u^{c}_j} = 
{\left(
\begin{array}{ccc}
\bar{h}_{1} +  
&-&-\\
\bar{h}_{2} (S_{1} S_{2} + S_{3} S_{4} + S_{5}S_{6})+ 
&&\\
-&
\bar{h}_{2}+ 
&-\\
&
  \bar{h}_{1} (S_{1} S_{2} + S_{3} S_{4} + S_{5}S_{6})+
&\\
&
 \bar{h}_{3} S_{7} \bar{S}_{8}
+ \bar{h}_{4} S_{9} \bar{\Phi}_{12} 
&\\
-&-&
\bar{h}_{3} + 
\\
&&
\bar{h}_{2}   S_{8}  \bar{S}_{7} + 
\bar{h}_{4}  S_9 \bar{\Phi}_{31}\\
\end{array} \right )}&&\nolabel
\\
\label{miju}
\eeqn
\vskip 1.0truecm

\no Table C.6 Possible Down-Quark and Electron Mass Matrix to 6$^{\rm th}$ Order
\vskip 0.3truecm
\beqn
M_{Q_i, d^{c}_j} = M_{L_i, e^{c}_j} = {\left(
\begin{array}{ccc}
h_{4}  \bS_{8} ( \bS_1 \bS_2 +\bS_3 \bS_4 +\bS_5 \bS_6)
&-&-\\  
&&\\
-& h_4\bar{S}_8 + & -\\
 & h_{3}  S_{9}  \bS_{7} \bP_{12}&\\
 & h_4(F_1 \bF_1 + K_4 \bK_4 + \sum_{i=1}^{9} S_i \bS_i &\\
& + \P_{12}\bP_{12} +\P_{3}\P_{3}) &\\
&&\\
-&-& h_4\bar{S}_7 \Phi_{2}\\
\end{array} \right )}\nolabel\\
\label{mijd2}
\eeqn
\vfill\newpage

\section{$D$- and Stringent $F$-Flat Directions Towards Optical Unification}

\no Table D.1 Classes of $D$- and $F$-Flat Directions and Related Massive Exotic Doublets. 
\begin{flushleft}
\begin{tabular}{|r |l | l |}
\hline
Class & massive pairs & orders of mass terms\\
\hline
 1-1   & $A_5 \bA_1$                            & 3 \\              
 1-2   & $A_7 \bA_6$                            & 7    \\           
 1-3   & $A_3 \bA_4$                            & 7    \\           
 2-1   & $A_5 (a_3 \bA_1 + a_7\bA_7)$, $A_2 \bA_4$
                                                & 3, 7    \\        
 3-1   & $A_4 \bA_2$, $A_7 \bA_5$, $A_6 \bA_1$  & 7, 7, 7 \\        
 3-2   & $(a_3 A_5 + a_7 A_6)\bA_1$, $A_2 \bA_4$, $A_5 \bA_7$
                                                & 3, 7, 7, 7\\      
 5-1   & $A_2 \bA_4$, $A_5 \bA_7$, $A_4 \bA_2$, $A_7 \bA_5$, $A_6 \bA_1$ 
                                                & 7, 7, 7, 7, 7\\   
 5-2   & $(a_3 A_5 + a_7 A_6) \bA_1$, $A_2 \bA_4$, $A_5 \bA_7$, $A_4 \bA_2$, $A_7 \bA_5$
                                                & 3, 7, 7, 7, 7\\   
\hline
\end{tabular}
\end{flushleft}
\newpage


\no Table D.2: $A_i\bA_j$ or $A_i A_j$ Mass-Generating Stringent Flat Directions 
(To At Least 17$^{\rm th}$ Order).\\

\no In Tables D.2-D.4, the first column specifies the VEV class, with the first component of the 
class designation indicating the number of independent pairs of massive exotic doublets
($A A$ or $A A$) produced, the second component distinguishing the mass combinations, and the
third component in Tables D.2-3 identifying a given flat direction.
The second column specifies the number of field VEVs, 
the third column specifies the order at which $F$-flatness is broken unless
self-cancellation via non-Abelian VEVs is induced
($\infty$ indicates $F$-flatness to all finite orders), 
the third column specifies the normalized anomalous charge, 
and the remaining columns specify the ratios of  
the norm-squared components of the field VEVs. 
($a_3$ and  $a_7$ denote varying normalized coefficients of mass eigenstate
components.) Note that none of these directions contain hidden sector $SU(5)$-charged fields. 
\begin{footnotesize}
\begin{flushleft}
\begin{tabular}{|l |r|r|r|rrrrrrrrrrrr|}
\hline
class
&$\# v$ 
& $F$-flat
& ${Q^{(A)}}'$& VEVs &&&&&&&&&&&\\
&&&                                                                                                    
& $S_9$ & $S_7$ & $S_8$ & $\bP_{12}$ & $\bP_{23}$ & $\bP_{31}$ & $S_1$ & $S_2$ & $S_3$ & $S_4$ & $S_5$ & $S_6$\\ 
&&& 
&$F_2$ & $\bF^{'}_3$ & $K_4$ & $K_1$ & $K_2$ & $K_3$ & $\bK^{'}_1$ & $\bK^{'}_2$ & $\bK^{'}_3$& $N_1$ & $N_2$& $N_3$\\
\hline
1-1&   8&$\infty$& -6&  24&  0&-18&  0&  0& 12&  0& -3&  0&  3&  0&  0\\&&&&  0&  0& -6&  6&  0&  0&  0&  0&  0&  0&  6&  0\\
   &   8&       &  -9&  33&  0&-27&  0&  0& 15&  0& -3&  0&  0&  0& -3\\&&&&  0&  0& -6&  0&  0&  6&  0&  0&  0&  0&  6&  0\\
   &   8&       &  -9&  33&  0&-27&  0&  0& 18&  0& -3& -3&  0&  0&  0\\&&&&  0&  0& -6&  6&  0&  0&  0&  0&  0&  0&  6&  0\\
   &   8&       &  -6&  24&  0&-18&  0&  0&  9&  3&  0&  0&  0&  0& -3\\&&&&  0&  0& -6&  0&  0&  6&  0&  0&  0&  0&  6&  0\\
   &   9&       &  -7&  27&  0&-21&  0&  0& 11&  0& -3&  0&  0&  2& -1\\&&&&  0&  0& -6&  0&  0&  6&  0&  0&  0&  0&  6&  0\\
   &   9&       &  -4&  18&  0&-12&  0&  0& 10&  0& -4&  0&  0&  0& -2\\&&&&  0&  0& -6&  0&  6&  0&  0&  0&  0&  0&  2&  4\\
   &   9&       &  -2&   9&  0& -6&  0&  0&  3&  0& -1&  0&  0&  2&  0\\&&&&  0&  0& -3&  0&  0&  3&  0&  0&  0&  0&  2&  1\\
   &   9&       & -12&  48&  0&-36&  0&  0& 24&  0& -9&  0&  3&  0&  0\\&&&&  0&  0&-12&  6&  6&  0&  0&  0&  0&  0& 12&  0\\
   &   9&       &  -2&  12&  0&-10&  0&  0&  8&  0& -1&  0&  1&  0&  0\\&&&&  0&  0& -2&  6&  0&  0&  4&  0&  0&  0&  2&  0\\
   &   9&       & -22&  72&-16&-50&  0&  0& 28&  0& -3&  0&  3&  0&  0\\&&&&  0&  0& -6&  6&  0&  0&  0&  0&  0&  0&  6&  0\\
   &   9&       &  -8&  30&  0&-24&  0&  0& 16&  0& -3& -2&  1&  0&  0\\&&&&  0&  0& -6&  6&  0&  0&  0&  0&  0&  0&  6&  0\\
   &   9&       &  -5&  21&  0&-15&  0&  0& 10&  1& -2&  0&  3&  0&  0\\&&&&  0&  0& -6&  6&  0&  0&  0&  0&  0&  0&  6&  0\\
   &   9&       & -15&  57&  0&-45&  0&  0& 27&  0& -9&  0&  0&  0& -3\\&&&&  0&  0&-12&  0&  6&  6&  0&  0&  0&  0& 12&  0\\
   &   9&       &  -1&   9&  0& -7&  0&  0&  3&  0&  1&  0&  0&  0& -3\\&&&&  0&  0& -2&  0&  0&  6&  0&  4&  0&  0&  2&  0\\
   &   9&       &  -5&  21&  0&-17&  0&  0&  9&  0& -1&  0&  0&  0& -3\\&&&&  0&  0& -4&  0&  0&  6&  0&  2&  0&  0&  4&  0\\
   &   9&       & -22&  72&-16&-50&  0&  0& 25&  0& -3&  0&  0&  3&  0\\&&&&  0&  0& -6&  0&  0&  6&  0&  0&  0&  0&  6&  0\\
\hline
\end{tabular}
\end{flushleft}
\end{footnotesize}
\newpage

\no Table D.2 continued:
\begin{footnotesize}
\begin{flushleft}
\begin{tabular}{|l |r|r|r|rrrrrrrrrrrr|}
\hline
class 
&$\# v$ 
& $F$-flat
& ${Q^{(A)}}'$& VEVs &&&&&&&&&&&\\
&&&                                                                                                    
& $S_9$ & $S_7$ & $S_8$ & $\bP_{12}$ & $\bP_{23}$ & $\bP_{31}$ & $S_1$ & $S_2$ & $S_3$ & $S_4$ & $S_5$ & $S_6$\\ 
&&& 
&$F_2$ & $\bF^{'}_3$ & $K_4$ & $K_1$ & $K_2$ & $K_3$ & $\bK^{'}_1$ & $\bK^{'}_2$ & $\bK^{'}_3$& $N_1$ & $N_2$& $N_3$\\
\hline
   &   9&$\infty$&-22&  72&-13&-53&  0&  0& 28&  0& -3&  0&  0&  0& -3\\&&&&  0&  0& -6&  0&  0&  6&  0&  0&  0&  0&  6&  0\\
   &   9&       &  -5&  21&  0&-15&  0&  0&  7&  1& -2&  0&  0&  3&  0\\&&&&  0&  0& -6&  0&  0&  6&  0&  0&  0&  0&  6&  0\\
   &   9&       &  -8&  30&  0&-24&  0&  0& 13&  1& -2&  0&  0&  0& -3\\&&&&  0&  0& -6&  0&  0&  6&  0&  0&  0&  0&  6&  0\\
   &   9&       &  -4&  18&  0&-12&  0&  0&  5&  3&  0&  0&  0&  2& -1\\&&&&  0&  0& -6&  0&  0&  6&  0&  0&  0&  0&  6&  0\\
   &  10&       &  -7&  27&  0&-21&  0&  0& 11&  0& -2&  0&  1&  0& -3\\&&&&  0&  0& -6&  0&  0&  6&  0&  0&  0&  2&  4&  0\\
   &  10&       &  -4&  18&  0&-12&  0&  0& 10&  0& -1&  0&  3&  0& -2\\&&&&  0&  0& -6&  6&  0&  0&  0&  0&  0&  0&  2&  4\\
   &  10&       &  -3&  12&  0& -9&  0&  0&  5&  0& -1&  0&  0&  1& -1\\&&&&  0&  0& -3&  0&  0&  3&  0&  0&  0&  0&  2&  1\\
   &  10&       & -15&  57&  0&-45&  0&  0& 27&  0& -6&  0&  3&  0& -3\\&&&&  0&  0&-12&  6&  0&  6&  0&  0&  0&  0& 12&  0\\
   &  10&       &  -5&  21&  0&-19&  0&  0& 12&  0& -1& -3&  0& -2&  0\\&&&&  0&  0& -2&  6&  0&  0&  0&  0&  4&  0&  2&  0\\
   &  10&       & -13&  51&  0&-39&  0&  0& 23&  0& -9&  0&  0&  2& -1\\&&&&  0&  0&-12&  0&  6&  6&  0&  0&  0&  0& 12&  0\\
   &  10&       &  -5&  21&  0&-19&  0&  0& 11&  0& -1&  0& -2&  0& -3\\&&&&  0&  0& -2&  0&  0&  6&  4&  0&  0&  0&  2&  0\\
   &  10&       & -22&  72&-15&-51&  0&  0& 26&  0& -3&  0&  0&  2& -1\\&&&&  0&  0& -6&  0&  0&  6&  0&  0&  0&  0&  6&  0\\
   &  10&       &  -6&  24&  0&-18&  0&  0&  9&  1& -2&  0&  0&  2& -1\\&&&&  0&  0& -6&  0&  0&  6&  0&  0&  0&  0&  6&  0\\
   &  11&       & -13&  51&  0&-39&  0&  0& 23&  0& -6&  0&  3&  2& -1\\&&&&  0&  0&-12&  6&  0&  6&  0&  0&  0&  0& 12&  0\\
1-6&  10&       &  -1&   9&  0& -5&  0&  0&  1&  0&  1&  0&  2&  0& -3\\&&&&  0&  0& -4&  0&  0&  6&  0&  2&  0&  4&  0&  0\\
   &  10&       &  -1&  15&  0& -7&  0&  0&  0&  0&  2&  0&  4&  1& -5\\&&&&  0&  0& -8&  0&  0& 12&  0&  4&  0&  8&  0&  0\\
   &  11&       &  -3&  75&  0&-39&  0&  0&  3&  0& 15&  0& 18&  0&-33\\&&&& 30& 30&-36&  0&  0& 36&  0&  0&  0& 36&  0&  0\\
   &  11&       &  -3&  81&  0&-39&  0&  0&  0&  0& 15&  0& 21&  3&-33\\&&&& 30& 30&-42&  0&  0& 42&  0&  0&  0& 42&  0&  0\\
1-7&  10&       &  -3&  21&  0&-11&  0&  0& 13&  0&  1&  0&  6&  0& -5\\&&&&  0&  0&-10& 12&  0&  0&  0&  2&  0&  0&  0& 10\\
\hline
\end{tabular}
\end{flushleft}
\end{footnotesize}
\newpage

\def\phte{\phantom{2-8}}
\def\phtt{\phantom{3-6}}
\no Table D.2 continued:
\begin{footnotesize}
\begin{flushleft}
\begin{tabular}{|l |r|r|r|rrrrrrrrrrrr|}
\hline
class \&
&$\# v$ 
& $F$-flat
& ${Q^{(A)}}'$& VEVs &&&&&&&&&&&\\
id $\#$&&&                                                                                                    
& $S_9$ & $S_7$ & $S_8$ & $\bP_{12}$ & $\bP_{23}$ & $\bP_{31}$ & $S_1$ & $S_2$ & $S_3$ & $S_4$ & $S_5$ & $S_6$\\ 
&&& 
&$F_2$ & $\bF^{'}_3$ & $K_4$ & $K_1$ & $K_2$ & $K_3$ & $\bK^{'}_1$ & $\bK^{'}_2$ & $\bK^{'}_3$& $N_1$ & $N_2$& $N_3$\\
\hline
2-1.1       &  10&$\infty$&  -3&  15&  0&-11&  0&  0&  7&  0& -2&  0&  3&  0&  1\\&&&&  0&  0& -4&  6&  0&  0&  0&  0&  2&  0&  4&  0\\
{\phte}.2   &  10&       &  -1&   9&  0& -5&  0&  0&  3&  2&  0&  0&  3&  0&  1\\&&&&  0&  0& -4&  6&  0&  0&  0&  0&  2&  0&  4&  0\\
{\phte}.3   &  11&       & -22&  72&-22&-48&  0&  0& 24&  0& -1&  0&  3&  0&  2\\&&&&  0&  0& -2&  6&  0&  0&  0&  0&  4&  0&  2&  0\\
{\phte}.4   &  11&       &  -1&   9&  0& -7&  0&  0&  5&  0& -1&  0&  2&  0&  1\\&&&&  0&  0& -2&  6&  0&  0&  2&  0&  2&  0&  2&  0\\
3-1.1       &  10&       &  -1&   9&  0& -5&  0&  0&  3&  0& -3& -2&  0&  0&  1\\&&&&  0&  0& -4&  0&  6&  0&  0&  0&  2&  4&  0&  0\\
\hline
\end{tabular}
\end{flushleft}
\end{footnotesize}

\vskip 0.5truecm

\no Table D.3: $A_i\bA_j$ or $A_i A_j$ Mass-Generating   
Stringent $F$-flat Directions (To At Least 17$^{\rm th}$ Order) Through Non-Abelian
Self-Cancellation.

\begin{footnotesize}
\begin{flushleft}
\begin{tabular}{|l|r|r|r|rrrrrrrrrrrr|}\hline
class \&
&$\# v$ 
& $F$-flat
& ${Q^{(A)}}'$& VEVs &&&&&&&&&&&\\
id $\#$&&&
& $S_9$ & $S_7$ & $S_8$ & $\bP_{12}$ & $\bP_{23}$ & $\bP_{31}$ 
& $S_1$ & $S_2$ & $S_3$ & $S_4$ & $S_5$ & $S_6$\\ 
&&& 
&$F_2$ & $\bF^{'}_3$ & $K_4$ & $K_1$ & $K_2$ & $K_3$ & $\bK^{'}_1$ & $\bK^{'}_2$ & $\bK^{'}_3$& $N_1$ & $N_2$& $N_3$\\
\hline
3-2.1     &  11& $>17$ (16) &   -3&  21&  0&-17&  0&  0& 13&  0& -5&  0&  0&  0&  1\\&&&& 0&  0& -4&  6&  6&  0&  6&  0&  2&  0&  4&  0\\
{\phtt}.2 &  11& $>17$ (16) &   -2&  18&  0&-16&  0&  0& 12&  0&  0& -2&  0&  0&  1\\&&&& 0&  0& -2&  6&  6&  0&  6&  0&  4&  0&  2&  0\\ 
{\phtt}.3 &  11& $>17$ (16) &   -3&  21&  0&-17&  0&  0& 13&  0&  0&  0&  1&  0&  1\\&&&& 0&  0& -4&  6&  6&  0&  6&  0&  2&  0&  4&  0\\ 
{\phtt}.4 &  12& $>17$ (16) &   -1&  15&  0&-13&  0&  0& 10&  0&  0& -1&  1&  0&  1\\&&&& 0&  0& -2&  6&  6&  0&  6&  0&  4&  0&  2&  0\\ 
\hline
\end{tabular}
\end{flushleft}
\end{footnotesize}

\begin{footnotesize}
\begin{flushleft}
\begin{tabular}{|l|l|}
\hline
class \& id $\#$ & constraints for $F$-flatness \\
\hline
3-3.1 &$ <K_2\cdot \bK^{'}_3> = 0$\\
3-3.2 &$ <K_2\cdot \bK^{'}_3> = 0$\\
3-3.3 &$ <K_2\cdot \bK^{'}_3> = 0$\\
3-3.4 &$ <K_2\cdot \bK^{'}_3> = 0$\\
\hline
\end{tabular}
\end{flushleft}
\end{footnotesize}
\hfill\vfill
\newpage

\no Table D.4: Directions Generating 5 Pairs of $A_i\bA_j$ or $A_i A_j$ Masses, but with  
$F$-Flatness Breaking Below 17$^{\rm th}$ Order\\
\no (Some of these directions have $F$-term self-cancellation below 11$^{\rm th}$ Order.) 
\vskip 0.01truecm
\begin{footnotesize}
\begin{flushleft}
\begin{tabular}{|l|r|r|r|rrrrrrrrrrrr|}\hline
class 
&$\# v$ 
& $F$-flat
& ${Q^{(A)}}'$& VEVs &&&&&&&&&&&\\
&&&
& $S_9$ & $S_7$ & $S_8$ & $\bP_{12}$ & $\bP_{23}$ & $\bP_{31}$ 
& $S_1$ & $S_2$ & $S_3$ & $S_4$ & $S_5$ & $S_6$\\ 
&&& 
&$F_1$ & $\bF^{'}_4$ & $K_4$ & $K_1$ & $K_2$ & $K_3$ & $\bK^{'}_1$ & $\bK^{'}_2$ & $\bK^{'}_3$& $N_1$ & $N_2$& $N_3$\\
\hline
5-1 &  11& 11 &   -1&  15&  0&-11&  0&  0&  6&  0& -6& -2&  0& -3&  1\\&&&&  0&  0& -4&  0& 12&  0&  0&  0&  8&  4&  0&  0\\  
    &  11& 12 &   -1&  15&  0& -9&  0&  0&  5&  0& -6& -2&  1& -2&  1\\&&&&  0&  0& -6&  0& 12&  0&  0&  0&  6&  6&  0&  0\\  
    &  11& 13 &   -1&  45&  0&-11&  0&  0&  6&  0& -6& -2&  0& -3&  1\\&&&& 30& 30& -4&  0& 12&  0&  0&  0&  8& 34&  0&  0\\  
    &  11& 14 &   -1&  45&  0& -9&  0&  0&  5&  0& -6& -2&  1& -2&  1\\&&&& 30& 30& -6&  0& 12&  0&  0&  0&  6& 36&  0&  0\\  
5-2 &  11& 12 &   -1&  15&  0&-11&  0&  0&  6&  0& -7&  0&  1& -3&  1\\&&&&  0&  0& -4&  0& 12&  0&  0&  0&  8&  2&  2&  0\\  
    &  11& 14 &  -20& 114&  0&-82&  0&  0& 51&  0&-27&  0&-14& -1& 10\\&&&& 30& 30& -2&  0& 24&  0&  0&  0& 22&  2& 30&  0\\  
    &  11& 13 &   -1&  39&  0& -7&  0&  0&  4&  0& -4&  0&  0& -1&  1\\&&&& 30& 30& -2&  0&  6&  0&  0&  0&  4& 30&  2&  0\\  
    &  11& 14 &   -1&  45&  0&-11&  0&  0&  6&  0& -7&  0&  1& -3&  1\\&&&& 30& 30& -4&  0& 12&  0&  0&  0&  8& 32&  2&  0\\  
\hline
\end{tabular}
\end{flushleft}
\end{footnotesize}

\hfill\vfill
\newpage

\end{document}